\documentclass[11pt]{article}
\usepackage{amssymb}
\usepackage{graphicx}
\usepackage{epsfig}

\begin{document}
 \baselineskip 0.21in

\title{\textbf{Weak Stability of $\ell_1$-minimization Methods in Sparse Data Reconstruction}}
\author{YUN-BIN ZHAO\thanks{School of Mathematics, University of Birmingham,
Edgbaston, Birmingham B15 2TT,  United Kingdom ({\tt
zhaoyy@maths.bham.ac.uk}).  The research of this author was partially supported by
the Engineering and Physical Sciences Research Council (EPSRC) under
  grant \#EP/K00946X/1.},  HOUYUAN JIANG  \thanks{Judge Business School, University of Cambridge,  Trumpington Street, Cambridge CB2 1AG,  United Kingdom
({\tt h.jiang@jbs.cam.ac.uk}).},  and ZHI-QUAN LUO\thanks{Shenzhen Research Institute of Big Data,
 Chinese University of Hong Kong, Shenzhen, Guangdong Province, China ({\tt luozq@cuhk.edu.cn}). }
}

\date{(1st version December 2015, 2rd version May 2016, Revised April 2017)}

\maketitle

  \noindent \textbf{Abstract.}  As one of the most plausible convex optimization methods for sparse
data reconstruction,  $\ell_1$-minimization plays a fundamental role in the development
of sparse optimization theory. The stability   of this method has been addressed  in the literature under various assumptions such as restricted isometry property (RIP),  null space property (NSP), and mutual coherence. In this paper, we propose a unified means
to develop the so-called weak stability theory for $\ell_1$-minimization methods under
the condition called weak range space property of a transposed design matrix, which turns out to be a necessary and sufficient condition for the standard  $\ell_1$-minimization method to be weakly stable in sparse data reconstruction. The reconstruction error bounds
established in this paper are  measured by the so-called Robinson's constant.
 We also provide a unified weak stability result   for standard $\ell_1$-minimization under several existing compressed-sensing matrix properties. In particular, the weak stability  of $\ell_1$-minimization under the constant-free range space property of order $k$ of the  transposed design matrix is established for the first time in this paper. Different from the existing analysis, we utilize the classic
Hoffman's Lemma concerning the error bound of linear systems   as well as the Dudley's theorem concerning the polytope
approximation of the unit $\ell_2$-ball to show that $\ell_1$-minimization is robustly and weakly stable in recovering sparse data from inaccurate measurements.    \\



\newpage

\section{Introduction}

 Data might be contaminated by some form of
random noise and the measurements of data are
 subject to quantization error.
 Thus a huge effort in sparse data reconstruction is made to ensure
the reconstruction algorithms stable in the sense that reconstruction
errors stay under control when the measurements are slightly
inaccurate and when the data is not exactly sparse (see, e.g.,
\cite{BDE09, E10,
 EK12, FR13}).  One of the widely used reconstruction  models is
the $\ell_1$-minimization
\begin{equation} \label{BP-0} \min_ {x} \{\|x\|_1:  ~ \|Ax-y\|_p \leq \varepsilon \}, \end{equation}  where
$\|\cdot \|_p$ is the  $\ell_p$-norm with $p\geq 1$ ($p=1,2,\infty$ will be considered in this
paper).  In the above model, $A \in \mathbb{R}^{m\times n} ~(m<n) $
is a full-row-rank
  matrix  called a design or sensing matrix  which is a
collection of known or learned dictionaries,   $
y=A\widehat{x} +u $ is the acquired measurement vector for the  data
$\widehat{x} $ to be reconstructed, and $u$ represents the measurement
error with level $\|u\|_p\leq \varepsilon. $
 The size of $\varepsilon$ is closely tied with the noise power. In this paper,   the given data $(A, y,
\varepsilon)$ is referred to as the problem data of (\ref{BP-0}). When $\varepsilon =0$, (\ref{BP-0}) is reduced to the so-called standard $\ell_1$-minimization, i.e.,
$ \min  \{\|x\|_1:  ~  Ax= y  \}. $
The use of $\ell_1$-norm
to promote sparsity in data processing has actually a long history
(see, e.g., \cite{L65, TBM79, LF81, DL92, M96, CDS98, M99}), but  a
significant development of   theory and algorithms for sparse
data  reconstruction has been made only recently in the framework of
compressed sensing (see, e.g. \cite{DE03, CT05, CRT06b, DET06,  C08,
E10, EK12}).

Assume that an unknown vector, denoted by $\widehat{x} ,$
satisfies $\|A\widehat{x} -y\|_p \leq \varepsilon. $ In traditional compressed sensing setting, it is generally assumed that problem (1) admits a unique optimal solution, in which case it is interesting to know how close the unique solution   of
(\ref{BP-0}) to  $\widehat{x}.$ This leads to the
traditional stability analysis  for $\ell_1$-minimization methods.
The major results in this aspect have been
achieved by Donoho, Cand\`es, Romberg, Tao, and others (e.g.,
\cite{DET06, CT05, CRT06b, C08}).  However, from a mathematical point of view, we still need to understand the general stability (which is referred to as the weak stability in this paper) of a reconstruction model by taking into account the settings where the problem might possess multiple optimal solutions or the sensing matrix $A$ might admit a certain less restrictive property than existing assumptions. Moreover, the study of weak stability will  also provide a novel  stability result under existing stability conditions.  Let us first recall the notation of best $k$-term approximation before we introduce the   weak stability.  Let $k$ be an integer number and
define
$$\sigma_k(x)_1 := \inf_{z} \{\|x-z\|_1: \|z\|_0\leq k\},$$
where $x\in \mathbb{R}^n   $ and $\|z\|_0 $ denotes the number of nonzero entries of $z \in \mathbb{R}^n. $
$\sigma_k(x)_1 $ is called  the $\ell_1$-error of best $k$-term
approximation.  Let $x^*$ be an optimal solution of
 (\ref{BP-0}) with given  problem data $(A, y, \varepsilon).$  Problem  (\ref{BP-0})  is said to be weakly stable for
noise-free reconstruction ($\varepsilon =0$) if for any feasible vector $x$ of the problem, there is a solution $x^*$ of  (\ref{BP-0}) such that
\begin{equation} \label{S} \|x-x^*\| \leq C \sigma_k(x)_1,
\end{equation}   where $\|\cdot\|$ is
a norm and $C $ is a constant depending on the  problem data $(A, y).$    Problem (\ref{BP-0})  is said to be robustly and weakly stable for noisy reconstruction ($\varepsilon >0$) if for any feasible vector  $x$ of the problem, there is a solution $x^*$ of (\ref{BP-0}) such that
\begin{equation} \label{RS} \|x-x^*\| \leq  C_1 \sigma_k(x)_1 +C_2
\varepsilon,
\end{equation}   where $ C_1 $ and $  C_2 $ are constants determined by the problem data $(A, y, \varepsilon).$

 When the solution $x^*$ of    (\ref{BP-0}) is unique (for instance, when $\varepsilon=0$ and when the matrix $A$ admits the restricted isometry property (RIP) or   null space property (NSP), see Definition 2.1), the weak stability can be reduced to the normal stability  if   constants  $C$, $C_1$ and $C_2$ are often measured in terms of RIP or NSP constants. Cand\`es and Tao \cite{CT05,
CT06} introduced the notion of the
 RIP   with constant $\delta_{K},$  where $K$ is a certain integer
number, and they proved in \cite{CT06} that if $\delta_{2k}+
\delta_{3k} <1$,  all $k$-sparse vectors can be exactly reconstructed
via  standard $\ell_1$-minimization. Furthermore,
Cand\`es, Romberg and Tao
\cite{CRT06b} have   shown that the   stability of problem (\ref{BP-0}) with $p=2$ can be guaranteed
 if $\delta_{3k}+3 \delta_{4k} <2. $ This result was
improved to $\delta_{2k} < \sqrt{2}-1 $ in \cite{C08}, and  was
further improved by several researchers (see, e.g.,  \cite{FL09,   CWX10a, ML11, CWX13,FR13, AS14}).  Finally, Cai and Zhang \cite{CZ14} has
improved this bound to $\delta_{2k} < 1/\sqrt{2}.$

The  NSP  of order $k$ (see Definition 2.1) is a necessary
and sufficient condition for every $k$-sparse vector  to be exactly
reconstructed with standard $\ell_1$-minimization. This NSP property appeared in
\cite{DH01, DE03, GN03} and was formally called the null
space property by Cohen et al. \cite{CDD09}.  The NSP is strictly
weaker than the RIP (see, e.g., \cite{F14, CCW15}).
It was shown
  \cite{CDD09, S12, EK12, FR13,  CCW15} that the stable NSP or
robust NSP (which is a strengthened version of the NSP of order $k$) guarantees the stability of $\ell_1$-minimization. A typical feature of  RIP- and NSP-based
  stability results for $\ell_1$-minimization methods is that the coefficients $C, C_1 $ and
$C_2$ in (\ref{S}) and (\ref{RS}) are measured by the RIP constant,   stable NSP constant or the robust NSP constant.

The  range
space property (RSP) of order $k$ of $A^T$ (see Definition 2.1) was introduced in \cite{Z13}. This property is also
   a necessary and sufficient condition for recovering  every
   $k$-sparse vector with standard $\ell_1$-minimization.  So this property is equivalent to the NSP of order $k.$ If the RSP is only defined locally  at a specific vector $x^*, $ it is called the individual RSP of $A^T$ at $x^*,$  which is  a nonuniform recovery condition for a specific vector \cite{Z13}. A stability analysis at a specific vector for $\ell_1$-minimization has been carried out in \cite{ZYY16}, under an assumption equivalent to the individual RSP.   Note that   RSP of order $k$ of $A^T$ and NSP of order $k$   are
 constant-free conditions in the sense that their definitions do not involve any constant, unlike the stable or robust NSP of order $k.$
 Although the stability of  $\ell_1$-minimization methods has  been extensively studied  under various conditions in the literature,  the weak stability of these methods has not been properly established at present.
 In this paper, we consider a more relaxed constant-free condition  than   RSP of order $k $ of $A^T. $  We ask whether the weak stability of $\ell_1$-minimization methods can be developed under less restrictive constant-free matrix properties than the existing ones.

  We  note that  the optimal solution $x^*$ of  (\ref{BP-0}) is not determined by the problem data $A$ only.  Clearly,  $x^*$ is
jointly determined by all  problem data $(A, y, \varepsilon) $  of (\ref{BP-0}). Different measurement vector
$y$ and noise level $\varepsilon$ together with different choice of the
norm in (\ref{BP-0}) will affect the  optimal solution of  (\ref{BP-0}) as well. In other words, in addition to   $A,$   the  problem data $(y, \varepsilon)$   will also
 directly or indirectly affect the reconstruction ability and
stability of $\ell_1$-minimization methods.  Exploiting
adequate problem data  will levitate the dependence
on the matrix property, and might yield a weak stability result
under less restrictive  assumptions than existing conditions.

 The purpose of this paper is to establish such weak stability  results for
 $\ell_1$-minimization methods under a constant-free and mild matrix property. We  prove that the so-called
 \emph{weak range space property  of $A^T$} (see
Definition 2.2) is a desired  sufficient condition
for  many $\ell_1$-minimization methods to be weakly stable in sparse data reconstruction. We  show that this condition is also necessary for standard $\ell_1$-minimization  to be weakly stable for any given measurement vector  $y \in \{Ax: \|x\|_0\leq k\}.$
This property is
directly tied to and originated naturally from the fundamental Karush-Kuhn-Tucker (KKK) optimality
conditions for linear optimization.  It is well known that the
optimality conditions completely characterize the
 optimal solutions $x^*$ of   $\ell_1$-minimization through
  problem data no matter whether the  optimal solution of the problem is unique or not. We will demonstrate that
  the weak RSP of
 order $k$ of $A^T,$ together with a classic error bound of
 linear systems developed by Hoffman \cite{H52} and
Robinson \cite{R73}, provides an efficient way to develop the  weak  stability theory for $\ell_1$-minimization.
Existing RIP, NSP,  mutual coherence conditions and their variants  imply the weak
 RSP of $A^T,$
 and we  show that  each of these existing conditions implies the same reconstruction error bounds  in terms of the so-called
Robinson's constants depending on the problem data.
 Moreover, the weak stability of $\ell_1$-minimization
under the RSP of order $k$ of $A^T$ or NSP of order $k$ is immediately obtained for the
first time, as a special case of the general weak stability
results established in this paper.

This paper is organized as follows. In section 2, we give the
definitions of several key matrix properties and  recall
the Robinson's constant and  Hoffman's lemma. We also prove that the
weak RSP of order $k$ of $A^T$ is a necessary condition for   standard
$\ell_1$-minimization with measurements  $y\in \{Ax: \|x\|_0\leq k\} $ to be weakly stable in sparse data reconstruction. In section 3, we
characterize  the weak stability of standard $\ell_1$-minimization   under the weak RSP. In section 4, we show the robust  weak
stability of the $\ell_1$-minimization problem with linearly representable
constraints, i.e., $p=1$ and $p=\infty$ in (\ref{BP-0}). In
section 5, we prove  the  robust weak stability of quadratically
constrained $\ell_1$-minimization.

\emph{Notation.} Unless otherwise stated, the identity matrix of any
order will be denoted by $I $ and a vector of ones will be denoted
by $e.$  The nonnegative orthant in $\mathbb{R}^n$ will be denoted
by $\mathbb{R}^n_+.$ The set of $m\times n $ matrices is denoted by
$\mathbb{R}^{m\times n } .$ The $p$-norm of a vector is defined as
$\|x\|_p = \left(\sum_{i=1} ^n |x_i|^p \right)^{1/p}, $ where $ p\geq 1.$
In particular, when
  $p=\infty,$  the $p$-norm is reduced to   $\|x\|_\infty =\max_{1\leq i\leq n} |x_i|.$ The induced
matrix norm of $A$ is defined as $\|A\|_{p\to q} = \max_{\|x\|_p\leq
1} \|Ax\|_q.$ For a vector $x\in \mathbb{R}^n,$ $|x|$, $(x)^+$ and
$(x)^-$ denote the vectors in $\mathbb{R}^n$ with components $|x|_i
:= |x_i|$, $ [(x)^+]_i := \max \{x_i, 0\} $ and  $ [(x)^-]_i := \min
\{x_i, 0\}, i=1, \dots, n,$ respectively. Given a subset $S\subseteq
\{1, \dots, n\}$ and a vector $x\in \mathbb{R}^n ,$      we use
$|S|$ to denote the cardinality of  $S$, $\overline{S}$ to denote
the complement of $S,$ i.e.,  $\overline{S} =\{1, \dots , n\}
\backslash S,$ and  $x_S$ to denote the subvector of $x$ by deleting
the components $x_i$ with $i\notin S. $    For matrix $A,$  $A^T$
denotes the transpose of $A, $  ${\cal R} (A^T)$ the range space of
$A^T, $  and ${\cal N} (A) $ the null space of $A. $   For any
vectors $x, y \in \mathbb{R}^n$, $x\leq y $ means $x_i\leq y_i$ for
all $i=1, \dots, n.$ A vector $x$ is said to be $k$-sparse if it
admits at most $k$ nonzero entries,  i.e., $\| x\|_0 \leq k.$

\section{Weak RSP of order $k$ of $A^T$  and Robinson's constant}

In this section, we provide some notions and  facts that
will be used throughout the remainder of the paper.  Let us first
recall some important matrix properties that have been widely used in  sparse recovery framework.

\vskip 0.07in

\textbf{Definition 2.1.} (a) (RIP of order $2k$) \cite{CT05, C08}
The matrix $A$ is said to satisfy the restricted isometry property
of order $2k$ with constant $\delta_{2k} \in (0,1)$ if
$ (1-\delta_{2k}) \|x\|^2_2 \leq \|Ax\|^2_2
\leq (1+\delta_{2k})\|x\|^2_2$ holds for all $k$-sparse vector
$x\in \mathbb{R}^n. $

 (b)  (NSP of order $k$) \cite{CDD09, Zh13, FR13} The matrix $A$ is said to satisfy the null space
property
 of order $k$   if
$\|v_S\|_1 < \|v_{\overline{S}}\|_1
 $ holds for any $v\in {\cal N}(A) $ and any $S\subseteq \{1, \dots, n\}$ with $|S| \leq
k. $

(c)  (Stable NSP of order $k$) \cite{ CDD09, Zh13, FR13}   The matrix
$A$ is said to satisfy the stable null space property of order $k$
with constant $\rho \in (0, 1)$   if $\|v_S\|_1 \leq \rho
\|v_{\overline{S}}\|_1
 $ holds for any $v\in  {\cal N}(A) $ and any $S\subseteq \{1, \dots, n\}$ with $|S| \leq
k. $

(d) (Robust NSP of order $k$)  \cite{CDD09, FR13}  The matrix $A$ is
said to satisfy the robust null space property of order $k$ with
constants $\rho \in (0, 1)$ and $\tau >0$ if $\|v_S\|_1 \leq \rho
\|v_{\overline{S}}\|_1 +\tau \|A v\|
 $ holds for any $v\in \mathbb{R}^n$ and any $S\subseteq \{1, \dots, n\}$ with $|S| \leq
k. $

(e)  (RSP of order $k$ of $A^T$) \cite{Z13}   The matrix $A^T$ is said to
satisfy the range space property of order $k$ if for any disjoint
subsets $S_1, S_2$ of $\{1,\dots , n\}$ with $|S_1|+|S_2|\leq k
 $, there is a vector $\eta  \in  {\mathcal R}(A^T) $ satisfying
 that
$\eta_i=1\textrm{ for }i\in S_1, ~ \eta_i=-1\textrm{
for  }i\in S_2,  ~|\eta_i| < 1 \textrm{ for  } i\notin S_1\cup
S_2. $

\vskip 0.07in

  The notion (e) above arises  from the uniqueness
analysis for the solution of linear $\ell_1$-minimization. In fact,
for any given $\widehat{x}$, it is known that $\widehat{x}$ is the
unique solution to the problem  $\min\{ \|z\|_1: Az =A\widehat{x}\}
$ if and only if $A_{supp(\widehat{x})}$ (the submatrix of $A$ formed by deleting the columns corresponding to the indices not in $supp (\widehat{x}) =\{i: \widehat{x}_i \not =0\} $) has full column rank and
the following property holds:   there is a vector $ \eta \in {\cal
R}(A^T) $ such that $\eta_i=1$ for $\widehat{x}_i
>0,$ $\eta_i=-1$ for $\widehat{x}_i <0, $ and $ |\eta_i|<1$ for
$\widehat{x}_i=0.$
  The sufficiency of the above statement was shown in \cite{F04}, and the necessity of the above statement was first shown in \cite{P07}.
   This fact was also rediscovered and proved independently in \cite{GSH11, Z13, FR13, ZYC15}.
 However, this  uniqueness property  depends on the individual vector $\widehat{x},$ and thus it   is insufficient for the uniform reconstruction of all $k$-sparse vectors via $\ell_1$-minimization. To exactly reconstruct every $k$-sparse vector  with $\ell_1$-minimization, this individual property is strengthened to  the  RSP of order $k $ of $A^T$ in \cite{Z13} so that it is independent of any individual vector.
  Given a matrix $A\in \mathbb{R}^{m\times n},$ it
is shown in \cite{Z13} that every $k$-sparse vector $\widehat{x}
\in \mathbb{R}^n$ can be exactly reconstructed by the $\ell_1$-minimization method
\begin{equation}  \label{LPXX}  \min\{\|z\|_1:  Az =y:= A
\widehat{x}\}  \end{equation}   if and only if $A^T$ admits the RSP
of order $k.$  So the RSP of order $k$ of $A^T$ is a necessary and sufficient
condition for the uniform recovery of all $k$-sparse vectors, and
hence it is equivalent to the NSP of order $k.$ An advantage of the RSP concept is that it can be easily extended to sparse data reconstruction with more complex structure than (\ref{LPXX})
(see, e.g., \cite{Z14, ZX16}).    We now introduce the weak RSP of order $k$  which is a relaxation of the
RSP of order $k. $

\vskip 0.07in

 \textbf{Definition 2.2.} (Weak RSP of order $k$ of $A^T$)     The matrix $A^T$ is said to
satisfy the weak range space property of order $k$ if for any
disjoint subsets $S_1, S_2$ of $\{1,\dots, n\}$ with
$|S_1|+|S_2|\leq
 k$, there is a vector $\eta  \in {\mathcal R}(A^T) $ satisfying
 that
 \begin{equation}
\label{SET-A} \eta_i=1\textrm{ for }i\in S_1, ~ \eta_i=-1\textrm{
for  }i\in S_2,  ~|\eta_i| \leq 1 \textrm{ for  } i\notin S_1\cup
S_2. \end{equation}

 \vskip 0.07in

Different from the RSP of order $k,$  the inequality ``$ |\eta_i| \leq
1 $ for $i \notin S_1\cup S_2$" in Definition 2.2 is not required to
hold strictly. The weak RSP of order $k $ of $A^T$ is a strengthened
optimality condition for the individual  problem (\ref{LPXX}). In
fact, by the KKT optimality condition, $\widehat{x}$ is an optimal
solution of (\ref{LPXX}) if and only if there is a vector $\eta  \in
{\cal R}( A^T)
  $ satisfying  $ \eta_i =1 $  for $\widehat{x}_i>0,$   $\eta_i= -
1  $ for $\widehat{x}_i<0 ,$  and $ |\eta_i|\leq  1 $ otherwise. Define the specific pair of $(S_1, S_2)$ with $ S_1=\{i: \widehat{x} _i >0\} $ and $   S_2=\{i: \widehat{x} _i <0\}. $ The  KKT optimality condition implies that the condition (\ref{SET-A}) holds for such a specific pair $(S_1, S_2).$ This can be  called the  individual weak  RSP of $A^T$ at   $\widehat{x}  . $  If we expect that  every $k$-sparse vector $\widehat{x}$ is an optimal solution to the $\ell_1$-minimization problem with measurements $y= A\widehat{x},$  then condition (\ref{SET-A}) must hold for any disjoint subsets $(S_1, S_2)$ with $|S_1\cup S_2| \leq k $ in order to cover all possible cases of $k$-sparse vectors. This naturally   yields the matrix property described in Definition 2.2.

The RIP of order $2k$  with  $\delta_{2k} \leq 1/\sqrt{2} $
implies that every
$k$ sparse vector  can be exactly recovered by  $\ell_1$-minimization (e.g., \cite{CZ14}). Thus it implies the RSP of order $k$ of $A^T$ which is equivalent to the NSP of order
$k.$   We see that the recovery condition $\mu_1(k)+\mu_1(k-1) <1  $ presented in  \cite{T04} also implies the NSP of order $k $ (see, e.g., Theorem 5.15 in \cite{FR13}), where $\mu_1(k)$ is the so-called accumulative coherence defined as
$$ \mu_1(k) =\max_{i\in \{1, \dots, n\}} \max \left\{ \sum_{j\in S} | a^T_ia_j| :  ~ S\subseteq \{1, \dots, n\}, ~
|S| =k,  ~ i\notin S\right\}, $$ where $a_i, i=1,\dots, n$ are the
$\ell_2$-normalized columns of $A.$
  Thus we have the following relation:
$$  \left. \begin{array}{r} \textrm{ RIP  of order } 2k   \Rightarrow   \\
\textrm{Stable  NSP  of order } k   \Rightarrow \\
\textrm{ Robust  NSP  of order } k   \Rightarrow \\
 \mu_1(k)+\mu_1(k-1) <1  \Rightarrow
 \end{array} \right\}
 \begin{array} {lll} \textrm{ NSP of order } k  & \Leftrightarrow  & \textrm{ RSP of order } k\textrm { of } A^T \Rightarrow  \\
 &  & \textrm{ weak RSP of order } k\textrm{ of } A^T.
 \end{array} $$
The weak RSP
  is the mildest  one  among the above-mentioned matrix properties.
To see how mild such a condition is, let us first prove   that the weak RSP of order $k$ of $A^T$  is a necessary condition  for standard $\ell_1$-minimization with any given measurement vector $y \in \{Ax: \|x\|_0\leq k\}$ to be weakly stable in sparse data reconstruction.

\vskip 0.07in

\textbf{Theorem 2.3.} \emph{Let $A$ be  a given $m \times n  ~(m<n)$
matrix with $rank (A)=m.$ Suppose that for any given measurement vector $y\in \{Ax: \|x\|_0 \leq k\},$  the following holds:  For any $x\in \mathbb{R}^n$ satisfying
$Ax=y ,$ there is a solution
 $x^*$ of the problem $\min\{\|z\|_1: Az= y  \}$
 such that
 $ \|x-x^*\| \leq  C \sigma_k(x)_1, $
where  $\|\cdot\|$ is a norm and $C$ is a constant dependent on the problem data $(A, y).$ Then $A^T$ must satisfy the weak RSP of order $k.$}

\vskip 0.07in

\emph{Proof.} Assume that $(S_1, S_2)$  is an arbitrary pair of
disjoint subsets of $  \{1, \dots, n\}$ with $|S_1|+|S_2| \leq k. $
Under the assumption of the theorem, we now prove that there exists
a vector $\eta \in {\cal R}(A^T) $ satisfying (\ref{SET-A}).  Then,  by Definition 2.2, $A^T$ must admit the
weak RSP of order $k.$ Indeed, let $\widehat{x}$ be a $k$-sparse
vector in $\mathbb{R}^n$ such that
\begin{equation} \label{S1S2}  \{i: \widehat{x}_i>0\}= S_1,  ~ \{i: \widehat{x}_i<0\}= S_2. \end{equation}     Consider the
problem  (\ref{LPXX}), i.e., $  \min\{\|z\|_1:  Az =y:= A
\widehat{x}\} . $  By the assumption,  there is an
optimal solution $x^*$ to this problem such that
$ \|\widehat{x}- x^*\| \leq C \sigma_k(\widehat{x})_1 ,$
where $C $ depends on the problem data $(A, y). $ Since
$\widehat{x} $ is $k$-sparse, the right-hand side of the inequality
above is equal to zero, and hence $\widehat{x} = x^*. $ This,
together with (\ref{S1S2}), implies that  \begin{equation}
\label{SET-B}\{i: x^*_i >0\} = S_1, ~ \{i: x^*_i <0\} = S_2, ~
x^*_i=0 \textrm{ for all }  i\notin S_1\cup S_2.\end{equation}  Note
that $x^*$ is an optimal solution to the convex problem
(\ref{LPXX}). $x^*$  must satisfy the optimality condition, i.e.,
there exists a vector $u\in \mathbb{R}^m $ such that
$ A^T u \in \partial \|x^*\|_1, $ where
$ \partial \|x^*\|_1 $ is the subgradient of the $\ell_1$-norm at
$x^*,$ i.e.,    $$ \partial \|x^*\|_1 =\{v \in \mathbb{R}^n:  ~ v_i
=1 \textrm{ for } x^*_i > 0, ~ v_i =-1 \textrm{ for } x^*_i <0, ~
|v_i| \leq   1 \textrm{ otherwise} \}. $$  By setting $\eta = A^T u \in \partial \|x^*\|_1,
$ we immediately see that $ \eta_i=1$ for $x^*_i>0$, $ \eta_i=-1$
for $x^*_i< 0,$ and $ |\eta_i | \leq 1$ for   $x^*_i=0.$  This,
together with (\ref{SET-B}), implies that the vector $\eta =A^T u$
satisfies (\ref{SET-A}). Since $S_1$ and $ S_2$  are  arbitrary disjoint subsets of $\{1, \dots, n\}$ with $ |S_1|+|S_2| \leq k.$ Thus $A^T$ must satisfy the weak RSP of order $k.$  ~ $\Box$

\vskip 0.07in

In the next section, we  show that the converse of the above result is also valid (see Theorem 3.2 and Corollary 3.3 for details).
  We will use a classic error bound for linear systems established by Hoffman \cite{H52}.  Let us first
recall a constant introduced by Robinson \cite{R73}. Let $P \in
\mathbb{R}^{n_1\times q}$ and $ Q \in \mathbb{R}^{n_2\times q} $ be
two real matrices. Define a set $F\subseteq \mathbb{R}^{n_1+n_2} $
by
   $$F = \{(b,d):\textrm{ for some } z \in \mathbb{R}^q \textrm{ such that }  P z \leq b \textrm{ and }  Q  z =d\}.$$
Let $\|\cdot\|_\alpha$ and $ \|\cdot\|_\beta $ be norms on
$\mathbb{R}^q$ and $ \mathbb{R}^{n_1+n_2}, $ respectively.
Robinson \cite{R73} has shown that the quantity \begin{equation}
\label{HR-mu} \mu_{\alpha, \beta} (P, Q):= \max_{\|(b,d)\|_\beta
\leq 1, (b,d)\in F}
 \min_{z\in \mathbb{R}^q} \{\|z\|_\alpha:  ~ P z\leq b,  ~ Q z =d\}\end{equation}  is a finite real number. It has also been shown in \cite{R73} that the extreme value above is attained.   In this paper, we use $\alpha = \infty, $ in which case $\|x\|_\infty$ is a polyhedral norm in the sense that the closed unit ball $\{ x: \|x\|_\infty \leq 1\} $ is a polyhedron.   Define the optimal value of  the  internal minimization in (\ref{HR-mu}) as  $$ g (b,d) =\min\{\|z\|_\alpha: ~P z \leq b, ~ Q  z =d\}, ~~ (b,d) \in F.  $$ Then  $$\mu_{\alpha, \beta} (P, Q)  = \max_{(b,d) \in \widetilde{B}\cap F} g (b,d),$$  where $\widetilde{B} =\{(b,d): ~ \|(b,d)\|_\beta \leq 1\}  $ is the unit ball in $\mathbb{R}^{n_1\times n_2}. $   As pointed out in \cite{R73}, the function $g(b,d)$ is convex over $F $ if  $\|\cdot\|_\alpha $ is a polyhedral norm.  In this case,   $\mu_{\alpha, \beta} (P, Q)$ is the maximum of a convex function over the bounded set $\widetilde{B}\cap F.$

   Let $M'\in \mathbb{R}^{m \times q }$ and $ M'' \in \mathbb{R}^{\ell
\times q } $ be two given matrices.
     Consider $(P, Q)  $ of the form  $$P = \left[
\begin{array}{cc}
 I_N & 0 \\
   -I & 0 \\
    \end{array}
    \right] \in \mathbb{R}^{(|N|+m)\times (m+\ell)},  ~ Q = \left[
          \begin{array}{c}
     M' \\
      M''\\
     \end{array}
     \right]^T \in \mathbb{R}^{q \times (m+\ell)},$$ where $ N  $ is a subset of $ \{1, \dots, m\}$ and  $I_N$ is obtained from the
      $m\times m$ identity matrix $I$ by deleting the rows corresponding to indices not in $N.$
        Robinson \cite{R73}  defined the following constant:
       \begin{equation} \label{HR-constant} \sigma_{\alpha, \beta}
(M', M''):= \max_{N\subseteq \{1, \dots , m\} } \mu_{\alpha, \beta}
\left(\left[
                \begin{array}{cc}
     I_N & 0 \\
       -I & 0 \\
        \end{array}
         \right], \left[
          \begin{array}{c}
               M' \\
            M''\\
           \end{array}
         \right]^T \right).
   \end{equation}
 As shown in \cite{R73}, the well known Hoffman's Lemma \cite{H52} in terms of
  constant (\ref{HR-constant}) with $(\alpha,\beta) =(\infty, 2)  $ is stated  as follows.

 \vskip 0.07in

   \textbf{Lemma 2.4.} (Hoffman)  \emph{Let $M'\in \mathbb{R}^{m\times q}$ and $ M'' \in \mathbb{R}^{\ell\times q}$ be given matrices and ${\cal F}
= \{x\in \mathbb{R}^q : M'x \leq b, M'' x= d\}. $
 For any vector $x$ in $\mathbb{R}^q, $  there is a point $x^* \in  {\cal F}$
 with
  $$ \|x-x^*\|_2 \leq  \sigma_{\infty, 2}(M', M'')
   \left\| \left[
    \begin{array}{c}
     (M'x-b)^+ \\
      M''x-d  \\
      \end{array}
      \right] \right\|_1.  $$}

The constant $ \sigma_{\alpha, \beta}(M', M''), $ defined in (\ref{HR-constant}), is referred to as the \emph{Robinson's constant} determined by $(M',M'').$  Given the solution set ${\cal F} $ of a linear system,  Hoffman's error bound claims that the distance from a point in space to   ${\cal F}$ can be measured in terms of the Robinson's constant and the quantity of the linear system being violated at this point.

  In the remainder of the paper, we use   Lemma 2.4  to
develop a weak-stability theory for $\ell_1$-minimization problems.
The purpose of this study   is to estimate the distance between an   unknown vector (which is the target data to reconstruct) and the   solution of the $\ell_1$-minimization problem. Note that the solution set of a linear optimization problem is a polyhedron which can be represented as the solution set of a certain linear system by using the KKT optimality condition. From this observation, a recovery error bound  via $\ell_1$-minimization is similar to the Hoffman's error bound, although they are not completely the same since sparsity is also involved in sparse data reconstruction. However, this similarity or connection motivates one to use Hoffman's error bound combined with sparsity assumption to form a new analytic method for studying stability issues in sparse data reconstruction. This is different from the standard analytic methods in this area.

  Our analysis not only provides a new tool to the study of stability issues of $\ell_1$-minimization, but also makes it possible to go beyond the standard framework of methods (such as RIP and NSP based ones) in order to develop   stability results under mild conditions or in general settings.
 As we have pointed out, most existing conditions can be relaxed to the assumption made in this paper. Traditional recovery error bounds are often established in terms of RIP constant, stable or robust stable NSP constant or their variants. Our assumption is a constant-free condition in the sense that the definition of this condition does not involve any constant that is difficult to certify. Under the constant-free  weak RSP of $A^T$ discussed in this paper,  we use Robinson's constant  to express stability coefficients in reconstruction error bounds.  The  error bound established under this assumption can apply to a wide range of matrix conditions, leading to a somewhat unified version of error bounds for sparse data reconstruction (see, e.g., Corollary 3.5). This is different from a standard analysis, which often requires an assumption-to-assumption analysis and the resulting error bounds often depends on an assumed individual assumption.  Hoffman's Lemma and Robinson's constant   provide a new perspective and an efficient way to interpret the sparse-signal-recovery behavior of $\ell_1$-minimization methods.

\section{Weak stability of $\ell_1$-minimization in noise-free settings}

In this section, we consider the case where the nonadaptive
 measurements $y\in \mathbb{R}^m   $ are accurate,  i.e., $y= A \widehat{x},$
where $\widehat{x}\in \mathbb{R}^m $ is the sparse data to reconstruct. The
situation with inaccurate measurements  will be discussed in later
sections. Given a  matrix $A$ and the noiseless measurements $y,$
the compressed sensing theory
  indicates that if  $A$ admits some strong property,  the standard $\ell_1$-minimization
\begin{equation}\label{BP} \min\{\|x\|_1: Ax=y\} \end{equation} can exactly reconstruct
the sparse data $\widehat{x}$ in the sense that the unique solution
$x^*$ of (\ref{BP}) coincides with $\widehat{x}.$  In many
situations, however,  the data  $\widehat{x}$  is not exactly sparse and it can only be
 claimed that  $\widehat{x}$   is close to a sparse vector. In these
situations, it is important to know  whether the reconstruction is weakly stable.     In section 2, we
have shown that the weak RSP of order $k$ of $A^T$ is a necessary
condition for  standard $\ell_1$-minimization with any given measurements $y \in \{Ax: \|x\|_0\leq k\} $  to be weakly stable.  In this section, we further show that this condition  is also sufficient for the problem
to be weakly stable. Note that the problem (\ref{BP})  can be written   as
the linear program
\begin{equation} \label{LP1} \min_{(x,t)} \{e^T t: ~Ax=y,  ~  -x+t \geq 0, ~x+t \geq 0,   ~t\geq 0\},  \end{equation}
to which the dual problem is given as
\begin{equation}\label{dual}
  \max_{(w, u,v)}   \left\{y^Tw  :    ~ A^Tw-u+v  = 0, ~   u+v \leq e, ~ (u,v) \geq 0
  \right\}.
 \end{equation}
 Thus, by the   optimality conditions, the solution of (\ref{BP}) can be characterized as  follows.

\vskip 0.07in

\textbf{ Lemma 3.1.}   \emph{ $x^*$ is an optimal solution of
  (\ref{BP}) if and only if there exist vectors
$t^* , u^* , v^* \in \mathbb{R}^n_+    $ and $ w^* \in \mathbb{R}^m
$ such that $ (x^*, t^*, u^*, v^*, w^*) \in D ,$ where
 \begin{eqnarray} \label{HHH-1} &  D=\{ (x,t,u,v,w):  &   Ax=y,  ~ x \leq t,
 ~ -x \leq t, ~ A^Tw-u+v=0,  ~ u+v \leq e, \nonumber \\
 & &        y^Tw= e^T t, ~ (u,v,t) \geq 0\}. \end{eqnarray}
 Moreover, any $(x,t,u,v,w) \in D $ satisfies that $t=|x|.$
}

\vskip 0.07in

The first assertion follows directly from the  optimality conditions
of  (\ref{LP1}) and (\ref{dual}). The second assertion is implied
from (\ref{HHH-1})  and   can  be directly seen from (\ref{LP1}) as
well. In fact, $x^*$ is an optimal solution  of (\ref{BP}) if and
only if $x^*,$ together with $t^*=|x^*|,$ is an optimal solution of
(\ref{LP1}).  Note that (\ref{HHH-1}) is of the form
\begin{equation} \label{HHH2}  D = \{z= (x, t, u,v, w): ~ M'z
\leq b, ~  M'' z = d  \}, \end{equation} where $ b= (0, 0, e, 0, 0, 0) $ and $d= (y, 0, 0)$ and
\begin{equation} \label{M'M''} M'= \left(
    \begin{array}{ccccc}
    I & -I & 0 & 0 & 0 \\
     -I & -I & 0 & 0 & 0 \\
       0 & 0 & I & I & 0 \\
         0 & 0 & -I & 0 & 0 \\
           0 & 0 & 0 & -I & 0 \\
            0 & -I & 0 & 0 & 0 \\
             \end{array}
              \right), ~ M''=\left(
               \begin{array}{ccccc}
                A & 0 & 0 & 0 & 0 \\
                 0 & 0 & -I & I & A^T \\
                  0 & -e^T & 0 & 0 &  y^T \\
            \end{array}
         \right). \end{equation}
    In the remainder of the paper, we use $c, c_1, c_2$ to denote the following constants:
        \begin{equation} \label{cc1c2} c =\|(AA^T)^{-1} A\|_{\infty \to \infty}, ~ c_1 =\|(AA^T)^{-1} A\|_{\infty \to 1}, ~ c_2 =\|(AA^T)^{-1} A\|_{\infty \to 2}. \end{equation}
         We now prove the main
result of this section.

\vskip 0.07in

\textbf{Theorem 3.2.}    \emph{Let $ A \in \mathbb{R}^{m\times n} ~(m<n) $ be a given matrix with
$\textrm{rank}(A)=m,$ and let $y$ be any given vector in
$    \mathbb{R}^m. $ If
  $A^T$ satisfies the weak RSP of order $k, $
   then, for any $x \in \mathbb{R}^n , $
    there is an optimal solution $x^*$ of  (\ref {BP})
   such that \begin{equation}\label{DELTA}  \|x-x^*\|_2 \leq  \gamma
    \left\{2\sigma_k(x)_1 + (1+c)\|Ax-y\|_1 \right\}, \end{equation}
    where $ c   $ is a constant given in (\ref{cc1c2}), and $\gamma= \sigma_{\infty, 2} (M', M'')   $  is the Robinson's constant with $ (M', M'')$  given as (\ref{M'M''}).  In particular, if $x$ satisfies  $Ax=y,$
     then  there is an optimal  solution $x^*$ of (\ref{BP}) such that   \begin{equation} \label{error-1} \|x-x^*\|_2 \leq  2 \gamma
     \sigma_k(x)_1. \end{equation}}

\emph{Proof.}   Let $x$ be any given vector in $\mathbb{R}^n$ and
let $t=|x|. $ Let $S$ denote the support set of the $k$-largest
components of $|x|.$
 Let $S=S_+ \cup S_-, $
 where $S_+ =\{i\in S:  x_i> 0\} $ and $ S_- = \{i\in S: x_i <  0\}.$
 We now construct a vector $(\widetilde{u}, \widetilde{v}, \widetilde{w})$  such that it is
 a feasible point to the problem (\ref{dual}).
Since $A^T$ has the weak RSP  of order $k$, there exists a
 vector $\eta \in {\cal R}(A^T) $ such that
  $  A^T\widetilde{w}   = \eta $ for some $\widetilde{w} \in \mathbb{R}^m $  and $\eta $ satisfies
  that
 $$ \eta_i= 1\textrm{ for }i\in S_+,
 ~\eta_i=-1 \textrm{ for }i\in S_{-}, ~|\eta_i| \leq 1\textrm{ for } i \notin  S  =S_+\cup S_-,$$
 from which we see that $ ( A^T\widetilde{w} )_S= \eta_S= \textrm{sign} (x_S).$
We construct $ (\widetilde{u},\widetilde{v}) $ as follows:
$ \widetilde{u}_i= 1\textrm{ and } \widetilde{v}_i=0\textrm{ for
}i\in S_+ ; $   $ \widetilde{u}_i=0\textrm{ and  }\widetilde{v}_i=
1\textrm{ for }i\in S_-;    $
 $ \widetilde{u}_i = (|\eta_i| +\eta_i)/2 \textrm{ and  } \widetilde{v}_i =(|\eta_i| -\eta_i)/2 \textrm{ for all }  i\notin S. $
From this construction,  $(\widetilde{u}, \widetilde{v})
$ satisfies that $ (\widetilde{u},\widetilde{v})\geq 0,  ~
\widetilde{u}+\widetilde{v}\leq e  $ and $ A^T\widetilde{w} = \eta
 = \widetilde{u}-\widetilde{v} . $
Thus $(\widetilde{u}, \widetilde{v}, \widetilde{w}) $ is a feasible
vector to the  problem (\ref{dual}). We now estimate the
distance of $(x, t, \widetilde{u}, \widetilde{v},
\widetilde{w}) $ to the set $D$ given by (\ref{HHH-1}) which can be written as (\ref{HHH2}). By applying Lemma 2.4 to (\ref{HHH2}), for the point $ (x, t,
\widetilde{u},\widetilde{v},\widetilde{w} )$, where $t=|x|,$  there
exists a point $ (x^*,t^*, u^*, v^*, w^*) \in D  $ such that
\begin{equation} \label{AAA1} \left\| \left[
                                        \begin{array}{c}
                                          x \\
                                          t \\
                                           \widetilde{u} \\
                                            \widetilde{v}\\
                                          \widetilde{w} \\
                                        \end{array}
                                      \right]
 -  \left[
                                        \begin{array}{c}
                                          x^* \\
                                          t^* \\
                                           u^* \\
                                           v^*\\
                                          w^* \\
                                        \end{array}
                                      \right]  \right\|_2 \leq
\gamma \left \| \left[
\begin{array}{c}
(x-t)^+\\
  (-x- t)^+  \\
(\widetilde{u}+\widetilde{v}-e)^+ \\
    Ax-y \\
    A^T\widetilde{w}-\widetilde{u}+\widetilde{v} \\
e^T t- y^T\widetilde{w}  \\
(\vartheta)^-\\
\end{array}
\right] \right \|_1, \end{equation}
 where  $(\vartheta)^-$ denotes the vector $ ((\widetilde{u})^-,
    (\widetilde{v})^-,
    (t)^- ), $   and  $\gamma =\sigma_{\infty, 2} (M',M'')$ is the  Robinson's
 constant determined by  $(M',M'') $   given as (\ref{M'M''}).
By the choice of $(\widetilde{u},\widetilde{v},\widetilde{w})$ and
the fact $t=|x|,$  we have
 $$  (x-t)^+=(-x- t)^+ =0, ~( \widetilde{u}+\widetilde{v}-e)^+=0,
  ~  A^T\widetilde{w}-\widetilde{u}+\widetilde{v}=0,  ~  (\vartheta)^-=0 .  $$ Thus the inequality (\ref{AAA1}) is reduced to
 \begin{eqnarray}  \label {1818}  \|(x, t, \widetilde{u},\widetilde{v},\widetilde{w} ) -
 (x^*, t^*, u^*, v^*, w^*)\|_2  & \leq  & \gamma \left \| \left[
  \begin{array}{c}
    Ax-y \\
    e^T t- y^T\widetilde{w}  \\
\end{array}
\right] \right \|_1 . \end{eqnarray}
 Denote by $h=Ax-y. $  By the
choice of $(t, \widetilde{u},\widetilde{v},\widetilde{w}), $ we see
that $$e^T t- y^T\widetilde{w} =e^T|x|- (Ax-h)^T \widetilde{w}=
\|x\|_1-x^T (A^T\widetilde{w})+ h^T \widetilde{w}.$$ Substituting
this into (\ref{1818}) and noting that $$ \|x-x^*\|_2 \leq \|(x, t,
\widetilde{u},\widetilde{v},\widetilde{w}) - (x^*, t^*, u^*, v^*, w^*)\|_2, $$  we obtain
 \begin{equation} \label{88-99} \|x-x^*\|_2    \leq
  \gamma  \left\{ \|Ax-y\|_1+ \left|\|x\|_1-x^T (A^T\widetilde{w}) + h^T \widetilde{w} \right| \right\}
 . \end{equation}
   Note that $A$ has full row rank and $\|\eta\|_\infty \leq 1.$   From $ A^T \widetilde{w} = \eta, $ we see
    that \begin{equation} \label{ccc}  \|\widetilde{w}\|_\infty
= \|(AA^T)^{-1}A \eta\|_\infty \leq \|(AA^T)^{-1}A\|_{\infty\to
\infty} \|\eta\|_\infty  \leq  c, \end{equation}  where $c
  $ is a constant given in (\ref{cc1c2}).   Note that $$ (x_S)^T (A^T\widetilde{w})_S
  =(x_S)^T \eta_S= (x_S)^T \textrm{sign} (x_S) =\|x_S\|_1.  $$ Therefore,
\begin{eqnarray} \left|\|x\|_1-x^T (A^T\widetilde{w})+ h^T \widetilde{w} \right|  & = &
 \left|\|x\|_1-(x_S)^T (A^T\widetilde{w})_S-(x_{\overline{S}})^T(A^T\widetilde{w})_{\overline{S}}  +  h^T \widetilde{w} \right| \nonumber \\
 & = & \left|\|x\|_1-\|x_S\|_1-(x_{\overline{S}})^T(A^T\widetilde{w})_{\overline{S}} + h^T \widetilde{w} \right|  \nonumber   \\  &= &
  \left| \sigma_k(x)_1 - (x_{\overline{S}})^T(A^T\widetilde{w})_{\overline{S}} + h^T \widetilde{w}\right|
  \nonumber \\
  & \leq &  \sigma_k(x)_1 + \left| (x_{\overline{S}})^T(A^T\widetilde{w})_{\overline{S}}| + |h^T
  \widetilde{w}\right|   \nonumber   \\
  &\leq &  2\sigma_k(x)_1 + \|h\|_1 \cdot \|\widetilde{w}\|_\infty  \nonumber   \\
  & \leq &  2\sigma_k(x)_1 +  c \|Ax-y\|_1 ,   \label{1919} \end{eqnarray}
  where  the second
  inequality follows from the fact
$$ \left| (x_{\overline{S}})^T(A^T\widetilde{w})_{\overline{S}} \right| \leq
\| x_{\overline{S}}\|_1 \left\|(A^T\widetilde{w})_{\overline{S}}\right\|_\infty
   = \|x_{\overline{S}}\|_1 \|\eta_{\overline{S}} \|_\infty \leq   \| x_{\overline{S}}\|_1 =   \sigma_k(x)_1,  $$ and the final inequality follows from   (\ref{ccc}).   Substituting (\ref{1919})
    into (\ref{88-99}) yields the  estimate (\ref{DELTA}),
   as desired.
In particular, if $x$ is a solution to the underdetermined linear system  $Az=y,$
then (\ref{DELTA})
 is reduced  to    (\ref{error-1}). ~  $\Box $

 \vskip 0.07in

 Under the weak RSP of order $k $ of $A^T, $ Theorem 3.2 indicates that the
  standard $\ell_1$-minimization problem, i.e., problem (1) with $\varepsilon =0$,  is  weakly stable for any given $y\in    \mathbb{R}^m   (= \{Ax: x\in \mathbb{R}^n\}$ since $A$ is underdetermined with full row rank).  In particular, it is weakly stable for any given $y \in   \{Ax: \|x\|_0\leq k\}\subseteq \mathbb{R}^m .$     Theorem 2.3 indicates that if the standard $\ell_1$-minimization problem is weakly stable for any given $y\in \{Ax: \|x\|_0\leq k\},$ then $A^T$ must satisfy the weak RSP of order $k. $ Merging Theorems 2.3 and 3.2 immediately yields the following statement.

  \vskip 0.07in

 \textbf{ Corollary 3.3.}  \emph{Let $A\in \mathbb{R}^{m\times n} (m<n) $ be a matrix with $\textrm{rank} (A)=m.$   Then the standard $\ell_1$-minimization problem  $\min\{ \|x\|_1:  Ax=y\} $ is weakly stable in sparse data reconstruction for any given measurements $y\in \{Ax: \|x\|_0\leq k\} $ if and only if  $A^T$   satisfies the weak RSP of order $k  .$ }

\vskip 0.07in

Thus the weak RSP of $A^T$ is the mildest condition, which cannot be relaxed without damaging the weak stability of $\ell_1$-minimization problems.

\vskip 0.07in

 \textbf{Remark 3.4.} Uniform recovery requires that every $k$-sparse vector  can be  reconstructed by $\ell_1$-minimization. This means that every $k$-sparse vector is an optimal solution of  $\ell_1$-minimization. Then the classic KKT optimality condition naturally yields the matrix property of weak RSP of $A^T.$  Therefore,  no matter what (deterministic or random) design matrix $A$ is used, the weak RSP of order $k$ of $A^T$ is a fundamental property required for achieving the uniform recovery  with $\ell_1$-minimization as a decoding method.
 The existence of a matrix with  such a property   follows directly from that of RIP matrices.  We recall the following fact:    (Cand\'es, Tao, etc.) \emph{Let A be an $m\times n$ Gaussian or Bernoulli random matrix. Then there exists a universal constant $C>0$ such that the RIP constant of $A/\sqrt{m}$ satisfies $\delta_{2k}  \leq \xi $ (where $0< \xi< 1$) with probability at least $1-\epsilon$ provided $$ m \geq C \xi ^{-2} \left(k (1+\ln (n/k))+\ln (2 \epsilon^{-1})\right). $$} This fact was first shown by Cand\`es and Tao \cite{CT05} and it was improved later to the above statement by Cand\`es and  other researchers.  Taking $\xi =1/\sqrt{2}$, Cai et al. \cite{CZ14} have shown that the RIP of order $2k$ with constant $\delta_{2k} < 1/\sqrt{2}$ guarantees the uniform recovery of $k$-parse vectors via $\ell_1$-minimization method. Note that the  uniform recovery of $k$-parse vectors via $\ell_1$-minimization is equivalent to that $A^T$ satisfies the RSP of order $k$ (see \cite{Z13} for details), and hence $A^T$ satisfies the weak RSP of order $k.$ Combining these facts and taking  $\xi = 1/\sqrt{2},$ we immediately obtain the following statement:
  \emph{Let A be an $m\times n$ Gaussian or Bernoulli random matrix. Then there exists a universal constant $C>0$ such that $A^T/\sqrt{m}$ satisfies the weak RSP of order $k$   with probability at least $1-\epsilon$ provided \begin{equation} \label{mmmm}  m \geq 2C \left(k (1+\ln (n/k))+\ln (2 \epsilon^{-1})\right). \end{equation} }
  By Theorem 3.2, when $A^T$ satisfies the weak RSP of order $k$, the error bound (\ref{error-1}) always holds. Combining  Theorem 3.2 and the above statements yields the following fact:  \emph{Let A be an $m\times n$ ($m<n$) Gaussian or Bernoulli random matrix with full row rank, and let $y$ be a given vector in $\mathbb{R}^m.$  Then there exists a universal constant $C>0$ such that
    with probability at least $1-\epsilon, $ the standard $\ell_1$-minimization problem with matrix $A/\sqrt{m}$ is weakly stable, provided that (\ref{mmmm}) is satisfied.}

\vskip 0.07in

 From
Theorem 3.2, we obtain a unified stability result
for several existing matrix properties.

\vskip 0.07in

 \textbf{Corollary 3.5.} \emph{Let $(A, y)$ be given, where $y \in \mathbb{R}^m $  and $ A \in \mathbb{R}^{m\times n} ~ (m<n) $ with  $\textrm{rank}(A)=m.$     Suppose that $A$ admits one of
the following properties:}
\begin{itemize}
\item[(p1)] \emph{RIP of order $2k$ with constant $\delta_{2k} < 1/\sqrt{2}. $}

\item[(p2)]\emph{ $A$ is a matrix with $\ell_2$-normalized columns and
$\mu_1(k) + \mu_1(k-1) <1, $  where $\mu_1(k)$ is the accumulated
mutual coherence.}

\item[(p3)] \emph{The stable NSP of order $k$ with constant $0<\rho <1.$}

\item[(p4)]  \emph{The robust NSP of order $k$  with constant $ 0<\rho <1$
and $\tau >0.$ }

\item[(p5)] \emph{The NSP of order $k. $ }

\item[(p6)] \emph{The RSP of order $k $ of $A^T.$ }
\end{itemize}
\emph{Then, for any $x \in \mathbb{R}^n , $
    the optimal solution $x^*$ of  (\ref {BP})
   approximates $x$ with error $$  \|x-x^*\|_2 \leq
     2 \gamma\sigma_k(x)_1 +  \gamma(1+c)\|Ax-y\|_1 , $$ where $ c $ is a constant given in (\ref{cc1c2}) and $\gamma =
\sigma_{\infty, 2} (M', M'') $  is the
 Robinson's constant determined by (\ref{M'M''}). In particular,  for any $x$ with $Ax=y,$ the optimal solution
   $x^*$ of   (\ref{BP}) approximates $x$ with error
$ \|x-x^*\|_2 \leq  2 \gamma  \sigma_k(x)_1. $   }

 \vskip 0.07in

The above corollary follows directly from Theorem 3.2, since each of
the properties (p1)--(p6) implies the weak RSP of order $k $ of $A^T$ as well as  the uniqueness of the optimal solution $x^*$ of  (\ref{BP}). Corollary 3.5 is a unified weak stability result
in the sense that   every matrix property of (p1)--(p6)
  implies the same error bound in terms of the  Robinson's
constant.  The weak stability result of this type is new and established in this paper for the first time.

\section{Robust weak stability of linearly constrained models}
 In more realistic situations,
the measurements $y$ for the unknown sparse data $\widehat{x} \in \mathbb{R}^n$ are inaccurate, and thus $y=A \widehat{x} +u, $
where $u$ denotes the measurement error satisfying $\|u \| \leq
\varepsilon $ for some norm $\| \cdot \|$ and noise level
$\varepsilon
>0.$ Thus we consider the robust weak stability of  (\ref{BP-0}) with a known level $\varepsilon >0.$  In this
section, we focus on the following problems:
\begin{eqnarray}  & \min   \{\|x\|_1:  &   \|Ax-y\|_\infty \leq \varepsilon
\}, \label{l1-infty} \\
  & \min   \{ \|x\|_1:  &   \|Ax-y\|_1 \leq \varepsilon \},   \label{l1-l1}
\end{eqnarray} corresponding to $p=\infty $ and
$p=1  $  in (\ref{BP-0}), respectively.
 The case $p=2$ in
(\ref{BP-0})  will be treated separately in section 5. Problems
(\ref{l1-infty}) and (\ref{l1-l1}) are referred to as the $\ell_1$-minimization with $\ell_\infty$-norm  and $\ell_1$-norm
constraints, respectively. A common feature of
(\ref{l1-infty}) and (\ref{l1-l1}) is that their constraints can be
linearly represented. This structure makes it possible to extend the
approach in section 3   to
establish the robust weak stability of
(\ref{l1-infty}) and (\ref{l1-l1}).

\subsection{$\ell_1$-minimization  with $\ell_\infty$-norm constraint}
We first consider the   problem (\ref{l1-infty}), which can be written as \begin{equation} \label
{LP-infty}  \min_{ (x,t) } \left\{e^T t: ~ -x + t \geq 0, ~ x+ t \geq 0,  ~
t\geq 0, ~ -\varepsilon e \leq Ax-y \leq \varepsilon
e \right\}\end{equation}  to which the dual problem is given as
\begin{equation} \label{dual-l1}    \max_{ (u,v, w, w') } \left\{  (y-\varepsilon e )^Tw - (y+\varepsilon e)^T w':
   ~ A^T (w-w') = u-v,   ~ u+v\leq e, ~(u,v, w, w')\geq 0 \right\}.
\end{equation}
 Clearly, $x^*$ is an optimal solution of (\ref{l1-infty}) if and only if $(x^*, t^*) $ with
  $t^* =|x^*|  $ is an optimal solution of (\ref{LP-infty}).  By the optimality condition  of a linear program,
 we can immediately characterize  the solution set of (\ref{l1-infty})  as follows.

 \vskip 0.07in

\textbf{Lemma 4.1. } \emph{ $x^*$ is an optimal solution of
(\ref{l1-infty}) if and only if there exist vectors $t^*, u^*, v^* $
in $ \mathbb{R}^n_+   $ and $ w^* ,  w'^* $ in $ \mathbb{R}^m_+ $
such that $ (x^*, t^*, u^*, v^*, w^*, w'^*) \in {D^{(\infty)}} $
where
 \begin{equation} \label{GGG} \begin{array} {ll} {D}^{(\infty)}=\{(x,t,u,v,w,w'):
 &  - x +  t \geq 0, ~ x+ t \geq 0, ~ -\varepsilon e \leq Ax-y
\leq \varepsilon e,    \\
 &   A^T
(w-w') = u-v, ~ u+v\leq e, \\
&  e^T t = (y-\varepsilon e )^Tw - (y+\varepsilon e)^T w', \\  &   (t,u,v, w, w') \geq 0 \}.
\end{array}    \end{equation}
 Moreover, for any $(x,t,u,v,w,w') \in {D}^{(\infty)}, $ it must hold that $t=|x|.$
}

 \vskip 0.07in

The set ${D}^{(\infty)}$ can be written  as
\begin{equation} \label{D1} {D}^{(\infty)} = \{z=(x,t,u,v,w,w'):  ~ M^{(1)} z  \leq b^{(1)},  ~ M^{(2)}
z = b^{(2)} \}, \end{equation}  where $b^{(2)}=0$ and
\begin{equation} \label{MM-X}
 M^{(1)}= \left(
\begin{array}{cccccc}
             I & -I & 0 & 0 & 0 & 0 \\
             -I & -I & 0 & 0 & 0 & 0\\
             A & 0 & 0 & 0 & 0 &  0 \\
             -A & 0 & 0 & 0 & 0 & 0 \\
             0 & -I & 0 & 0 & 0 & 0 \\
             0 & 0 & I & I & 0 & 0 \\
             0 & 0 & 0 & 0 & -I_m & 0 \\
             0 & 0 & 0 & 0 & 0 & -I_m \\
             0 & 0 & -I & 0 & 0 & 0 \\
                0 & 0 &   & -I & 0 & 0 \\
           \end{array}
         \right),  ~~~ b^{(1)}= \left(
                        \begin{array}{c}
                          0 \\
                          0 \\
                          y+\varepsilon e \\
                          \varepsilon e -y \\
                          0 \\
                          e \\
                          0 \\
                          0 \\
                          0 \\
                        \end{array}
                      \right),
                      \end{equation}
 \begin{equation} \label{MM-Y}
         M^{(2)} = \left(
           \begin{array}{ccccccc}
             0 & 0 & -I & I & A^T & -A^T  \\
              0 &  e^T & 0  & 0 & -(y-\varepsilon e)^T &  ( y+\varepsilon e)^T  \\
               \end{array}
               \right), \end{equation}
               where $I $ and $I_m$ are $n$- and $m$-dimensional identity matrices, respectively.
We now show that the robust weak stability  of (\ref{l1-infty}) is guaranteed under the weak RSP of order
$k$ of $A^T.$

 \vskip 0.07in

\textbf{Theorem 4.2.}    \emph{Let the problem data $(A, y,
\varepsilon)$  of (\ref{l1-infty}) be given, where $\varepsilon>0,$ $ y\in \mathbb{R}^m$ and   $A\in
\mathbb{R}^{m\times n}~  (m<n) $ with $\textrm{rank} (A) =m.$
Let $A^T$ satisfy the weak RSP of order $k. $
   Then for any $x \in \mathbb{R}^n , $
    there is  an optimal solution $x^*$ of (\ref{l1-infty})
    such that $$ \|x-x^*\|_2 \leq
     \gamma_1
\left\{\|(Ax-y-\varepsilon e)^+\|_1+  \|(Ax-y+\varepsilon e)^-\|_1
+2\sigma_k(x)_1 + c_1 \varepsilon  + c_1\|Ax-y\|_\infty \right\}, $$
where $c_1  $ is the constant given in (\ref{cc1c2}) and $\gamma_1
=\sigma_{\infty,2}(M^{(1)},M^{(2)}) $
is the Robinson's constant determined by    $(M^{(1)},M^{(2)})$ given
in (\ref{MM-X}) and (\ref{MM-Y}).  In particular, for any $x$ with
$\|Ax-y\|_\infty \leq \varepsilon, $
      there is a solution $x^*$ of (\ref{l1-infty})
     such that  $$ \|x-x^*\|_2 \leq
     2\gamma_1 \{\sigma_k(x)_1 +   c_1 \varepsilon\}. $$}

\emph{Proof.}  For any given $x\in \mathbb{R}^n$, we   consider a
vector  $(t, u,v, w, w')$ satisfying the following properties:
$t=|x|$ and $(u,v, w,w')$ satisfies $  A^T (w-w') = u-v,   u+v\leq e
$ and $ (u,v,w,w') \geq 0, $  i.e.,  $(u,v, w,w')$ is a feasible
vector to  problem (\ref{dual-l1}).
  Note that the set (\ref{GGG}) can be written as (\ref{D1}).  For such a  vector $(x, t, u, v, w, w')$, applying Lemma 2.4 with $(M',M'')= (M^{(1)},
  M^{(2)})$ being given in (\ref{MM-X}) and (\ref{MM-Y}),
   there must exist  a vector
  $(x^*,t^*,u^*, v^*, w^*, w'^*) \in
  {D}^{(\infty)} $ such that
\begin{equation} \label {hoff} \left\|
   \left[
      \begin{array}{c}
        x \\
        t \\
        u \\
        v \\
        w \\
        w' \\
      \end{array}\right] - \left[
       \begin{array}{c}
        x^* \\
         t^* \\
        u^* \\
         v^* \\
        w^* \\
         w'^* \\
       \end{array}
     \right]
 \right\|_2 \leq  \gamma_1 \left\| \left[\begin{array}{c}
   (x-t)^+\\
   (-x- t)^+  \\
   (Ax-y-\varepsilon e)^+ \\
 (Ax-y+ \varepsilon e)^-\\
    A^T(w-w')-u+v \\
    ( u+v-e)^+ \\
      e^T t - (y-\varepsilon e )^Tw+(y+\varepsilon e)^T w' \\
      (\widehat{\vartheta})^-
            \end{array}
             \right] \right\|_1, \end{equation}
where $ (\widehat{\vartheta})^- $  is short for the vector $ ( (u)^-, (v)^-,
(t)^-, (w)^-, (w')^-),$   and $\gamma_1 =$$\sigma_{\infty, 2}
(M^{(1)}, M^{(2)}) $  is the Robinson's constant with $ (M^{(1)},
M^{(2)}) $ being given by (\ref{MM-X}) and (\ref{MM-Y}).  By the nonnegativity of $(u,v,t,w,w'), $  we see that $ (\widehat{\vartheta})^- =0. $ Since $t=|x|$ and $(
u,v, w,w') $ is feasible to problem (\ref{dual-l1}), we see that
$$ (x-t)^+=(-x- t)^+ =0, ~ A^T(w-w') -u+v=0, ~ (u+v-e)^+=0 . $$    Thus  the system
(\ref{hoff}) is reduced to
\begin{eqnarray*}
  \|(x, t, u,v,w,w')-
 (x^*,t^*,u^*, v^*, w^*, w'^*) \|_2
\leq   \gamma_1  \left\| \left[
  \begin{array}{c}
     (Ax-y-\varepsilon e)^+ \\
 (Ax-y+\varepsilon e)^-\\
      e^T t - (y-\varepsilon e )^Tw + (y+\varepsilon e)^T w' \\
\end{array}
\right] \right\|_1.    \label{S-Hoff}
      \end{eqnarray*}
  Let $\phi =y -Ax.$
   We see  that
\begin{eqnarray*}    e^T t - (y-\varepsilon e)^Tw+(y+ \varepsilon e)^T w'
& = & e^Tt-
y^T(w-w') +  \varepsilon e^T(w+w')  \nonumber \\
 & = & e^T|x|-  (Ax+\phi)^T(w-w') +  \varepsilon e^T(w+w')  \nonumber \\
& = & e^T|x|-  x^T A^T (w-w') - \phi^T(w-w') + \varepsilon
e^T(w+w').
\end{eqnarray*}
Merging the above two relations leads to
\begin{eqnarray} \label{INQ}
 & &    \|  (x, t, u,v,w,w') -
 (x^*,t^*,u^*, v^*, w^*, w'^*)\|_2
\nonumber \\
 &    &  \leq    \gamma_1 \{ \|(Ax-y-\varepsilon e)^+\|_1+  \|(Ax-y+\varepsilon
 e)^-\|_1  \nonumber \\
  &  & + \left| e^T|x|-  x^T A^T (w-w') - \phi^T(w-w') +  \varepsilon
e^T(w+w')\right|\}.
\end{eqnarray}
By the weak RSP of order $k$ of $A^T,$ we now construct a  specific
vector $(\widetilde{u},\widetilde{v},\widetilde{w},\widetilde{w}')$
which is feasible to  problem (\ref{dual-l1}). To this goal,
 let $S$ denote the support set of the $k$-largest components of
 $|x|.$
   Let $S_+ =\{i\in S:  x_i> 0\} $
 and $ S_- = \{i\in S: x_i <  0\}. $  Clearly, $S=S_+ \cup S_-. $  Since $A^T$ satisfies the weak RSP
 of order $k, $  there exists a vector $\eta \in {\cal R}(A^T) $ such that $  \eta=  A^Tg   $
 for some $g \in \mathbb{R}^m$ and $\eta$ satisfies the following conditions:
 $$ \eta_i= 1\textrm{ for }i\in S_+, ~ \eta_i=-1 \textrm{ for all }i\in S_{-}, ~ \textrm{ and }
 |\eta_1| \leq 1\textrm{ for  } i \in \overline{S} =\{1, \dots , n\}\setminus S.$$
Construct $(\widetilde{u},\widetilde{v})$  as follows:
$\widetilde{u}_i= 1\textrm{ and } \widetilde{v}_i=0\textrm{ for
}i\in S_+;  ~\widetilde{u}_i=0\textrm{ and  }\widetilde{v}_i=
1\textrm{ for }i\in S_-;  $   $ \widetilde{u}_i =(1+\eta_i)/2 $ and
$\widetilde{v}_i = (1-\eta_i)/2 \textrm{ for all }i\in \overline{S}.
$ By this construction, we see that $\widetilde{u}- \widetilde{v}
=\eta.$ Moreover, by setting $\widetilde{w}  =(g)^{+} $ and
$\widetilde{w}' =-(g)^{-}, $ we see that $ \widetilde{w}  \geq 0, ~
\widetilde{w} ' \geq 0,  ~ \widetilde{w} -\widetilde{w} ' =g .$ It is easy to see that the vector $
(\widetilde{u}, \widetilde{v}, \widetilde{w} , \widetilde{w} ' )$
specified as above satisfies the conditions
$$   \widetilde{u}+\widetilde{v}\leq e,
~ A^T (\widetilde{w} -\widetilde{w} ')  =
\widetilde{u}-\widetilde{v}, ~ (\widetilde{u},\widetilde{v},
\widetilde{w} , \widetilde{w} ') \geq 0
$$ which indicates that $(\widetilde{u}, \widetilde{v}, \widetilde{w} ,
\widetilde{w} ') $
  is a feasible vector to  problem (\ref{dual-l1}). Thus
it follows from (\ref{INQ}) that for the vector  $(x, t=|x|,
\widetilde{u}, \widetilde{v}, \widetilde{w} , \widetilde{w} ')$,
there is a point in ${D}^{(\infty)},$ denoted still by $(x^*,t^*,u^*, v^*,
w^*, w'^*)$, such that
\begin{eqnarray} \label{INQ-2}
 & &    \|  (x, t, \widetilde{u},\widetilde{v},\widetilde{w} ,\widetilde{w} ') -
 (x^*,t^*,u^*, v^*, w^*, w'^*) \|_2
\nonumber \\
 &    &  \leq    \gamma_1  \{ \|(Ax-y-\varepsilon e)^+\|_1+  \|(Ax-y+\varepsilon
 e)^-\|_1  \nonumber \\
  &  & + \left| e^T|x|-  x^T A^T (\widetilde{w} -\widetilde{w} ') - \phi^T(\widetilde{w} -\widetilde{w} ')+ \varepsilon
e^T(\widetilde{w} +\widetilde{w} ')\right|  \}.
\end{eqnarray}
By the construction of $ (\widetilde{u}, \widetilde{v},
\widetilde{w} , \widetilde{w} ') ,$ we see that  $[A^T(
\widetilde{w} - \widetilde{w} ' )]_S=(\widetilde{u}-\widetilde{v})_S
=\eta_S= \textrm{sign} (x_S). $  Thus
\begin{eqnarray}  \label{INQ-3}  & & \left| e^T|x|-  x^T A^T (\widetilde{w} -\widetilde{w} ')
- \phi^T(\widetilde{w} -\widetilde{w} ') +
 \varepsilon e^T(\widetilde{w} +\widetilde{w} ')  \right| \nonumber  \\
 & & = \left|\|x\|_1-(x_S)^T [A^T (\widetilde{w} -\widetilde{w} ')]_S
 -(x_{\overline{S}})^T[A^T (\widetilde{w} -\widetilde{w} ')]_{\overline{S}} - \phi^T(\widetilde{w} -\widetilde{w}
 ')+
\varepsilon e^T(\widetilde{w} +\widetilde{w} ') \right|  \nonumber  \\
 &  &  = \left|\|x\|_1-\|x_S\|_1   -(x_{\overline{S}})^T[A^T (\widetilde{w} -\widetilde{w} ')]_{\overline{S}}
 - \phi^T(\widetilde{w} -\widetilde{w} ') +
 \varepsilon e^T(\widetilde{w} +\widetilde{w} ') \right|  \nonumber  \\
& & \leq \sigma_k(x)_1 + |(x_{\overline{S}})^T[A^T (\widetilde{w}
-\widetilde{w} ')]_{\overline{S}}|+ | \phi^T(\widetilde{w}
-\widetilde{w} ')| + \varepsilon |e^T(\widetilde{w} +\widetilde{w}
') |
 \nonumber  \\
&  & \leq  \sigma_k(x)_1 + \|(x_{\overline{S}})\|_1 \|[A^T
(\widetilde{w} -\widetilde{w} ')]_{\overline{S}} \|_\infty + |\phi^T
g|+ \varepsilon|
e^T((g)^+-(g)^-)|  \nonumber  \\
& & \leq 2\sigma_k(x)_1 +  \|g\|_1 \|\phi\|_\infty +  \varepsilon
\|g\|_1,
\end{eqnarray}
where the last inequality follows from the fact $\|[A^T
(\widetilde{w}-\widetilde{w}')]_{\overline{S}}\|_\infty
=\|\eta_{\overline{S}}\|_\infty \leq 1.$ Since $A^T$ has full column
rank, it follows from $A^T g= \eta $ that $g=(AA^T)^{-1}A \eta, $ and
hence
\begin{equation}
\label{INQ-4} \|g\|_1 = \|(AA^T)^{-1}A \eta \|_1 \leq
 \|(AA^T)^{-1}A\|_{\infty \to 1} \| \eta \|_\infty \leq \|(AA^T)^{-1}A\|_{\infty \to
 1} = c_1.  \end{equation}
Merging   (\ref{INQ-2}), (\ref{INQ-3}) and
(\ref{INQ-4}) yields the bound
 \begin{eqnarray*}   \|x- x^* \|_2
 &    \leq &    \|(x, t, \widetilde{u},\widetilde{v},\widetilde{w},\widetilde{w}') -
(x^*,t^*,u^*, v^*, w^*, w'^*) \|_2
\\
 &  \leq  &    \gamma_1 \left\{\|(Ax-y-\varepsilon e)^+\|_1+
\|(Ax-y+\varepsilon
 e)^-\|_1 +  2\sigma_k(x)_1 +  \|g\|_1 (\|\phi \|_\infty + \varepsilon) \right\} \\
  & \leq  &   \gamma_1 \left\{\|(Ax-y-\varepsilon e)^+\|_1+
\|(Ax-y+\varepsilon
 e)^-\|_1 +  2\sigma_k(x)_1 +  c_1\|y-Ax\|_\infty + c_1\varepsilon \right\},
\end{eqnarray*}
as desired. In particular, when $x$ satisfies the constraint of
(\ref{l1-infty}), i.e., $\|y-Ax\|_\infty \leq \varepsilon,$ the
above estimate reduces to $ \|x- x^* \|_2    \leq     2 \gamma _1 \{
\sigma_k(x)_1 +  c_1\varepsilon \}. $  ~~ $\Box$

 \subsection{$\ell_1$-minimization with $\ell_1$-norm constraint}

We now show the robust weak stability  of   problem (\ref{l1-l1}). Note that
(\ref{l1-l1}) is equivalent to
\begin{equation}\label{l1-A}   \min_{(x,r)} \left\{\|x\|_1:  ~ |Ax-y|\leq r, ~ e^T r \leq \varepsilon, ~ r\in
\mathbb{R}^m_+ \right\}. \end{equation} It is evident that $x^*$ is an
optimal solution of  (\ref{l1-l1}) if and only if there is a vector
$r^*$   such that  $(x^*, r^*)$ is an
optimal solution of (\ref{l1-A}).
We may further write (\ref{l1-A}) as the linear program
\begin{equation} \label{LP-A} \min_{(x,t,r)}  \left\{ e^T t :
  ~ x \leq t, ~  -x\leq t, ~ t\geq 0, ~ Ax-y\leq r, -Ax+y \leq r,
  ~  e^T r \leq \varepsilon,  ~ r \geq 0 \right\}.
 \end{equation}
 The dual problem of (\ref{LP-A}) is given by
 \begin{eqnarray} \label{DLP-A} &\max & -y^T
 (v_3-v_4)-\varepsilon v_5  \nonumber \\
 & \textrm{s.t} & A^T (v_3-v_4) +v_1-v_2=0,
  ~ v_3+v_4 \leq v_5 e,
   ~ v_1+v_2\leq e,  \\
   & &  v_i\geq 0,  ~i =1, \dots, 5,  \nonumber
  \end{eqnarray}
where $v_1,  v_2\in \mathbb{R}^n_+, v_3, v_4\in \mathbb{R}^m_+,$ and
$ v_5\in \mathbb{R}_+. $
 By the optimality condition  of a linear program, the solution set of (\ref{l1-l1}) can be characterized as
 follows.

 \vskip 0.07in

\textbf{Lemma 4.3.} \emph{ $x^*$ is an optimal solution   of
(\ref{l1-l1}) if and only if there exist vectors $t^*, v_1^*, v_2^*
\in \mathbb{R}^n_+, v_3^*, v_4^*, r^* \in \mathbb{R}^m_+  $ and $v_5^*
\in \mathbb{R}_+$ such that $ (x^*, t^*, r^*, v_1^*, \dots , v_5^*)
\in {D}^{(1)} , $ where
 \begin{equation} \label{QQQ} \begin{array} {ll}  D^{(1)}=\{(x,t, r, v_1, \dots, v_5):
 &~  x\leq  t , ~ -x \leq t  , ~ Ax-r \leq y,   ~  -Ax-r \leq -y, \\
 & ~ e^T r \leq \varepsilon,
  ~ (r , t) \geq 0, ~  A^T (v_3-v_4) + v_1-v_2 =0, \\
  & ~ v_3+v_4- v_5 e\leq 0,
   ~ v_1+v_2\leq e,  \\
&~  e^T t = -y^T(v_3-v_4)-v_5 \varepsilon,  ~  v_i\geq 0,  ~i=1, \dots , 5 \}.
\end{array}  \end{equation}
 Moreover, for any $ (x,t, r, v_1, \dots , v_5) \in {D}^{(1)},  $ it must hold that
 $t=|x|.$ }

\vskip 0.07in

 In order to apply Lemma 2.4 in the proof of the next theorem, we  rewrite
  ${D}^{(1)} $  as
  \begin{equation} \label{M*M**} D^{(1)} =  \left\{z= (x,t, r, v_1, \dots, v_5):  ~ M^* z\leq b^* ,
  ~ M^{**} z = b^{**} \right\}, \end{equation}  where  $b^{**} =0, $
 $b^{*}$ is a vector  consisting  of $0, y, -y, e
 $ and $
 \varepsilon,$ The matrix $M^*$
  captures
  all  coefficients of the inequalities in (\ref{QQQ}),  and   $M^{**}$ is the matrix
  capturing all coefficients of the equalities in (\ref{QQQ}). The entries of $M^*$ and $ M^{**}$ are given by the problem data $ (A, y, \varepsilon).$  $M^*$ and $M^{**}$ are
 omitted here. We have the following stability result.

 \vskip 0.07in

\textbf{Theorem 4.4.}    \emph{Let the problem data $(A, y,
\varepsilon)$  of (\ref{l1-l1}) be given, where $\varepsilon >0,$ $y\in \mathbb{R}^m $ and  $A\in
\mathbb{R}^{m\times n}~  (m<n) $ with $\textrm{rank} (A) =m.  $
Let $A^T$ satisfy the weak RSP of order $k.$
   Then for any $x \in \mathbb{R}^n , $
    there is  an optimal solution $x^*$ of (\ref{l1-l1})
     such that  $$ \|x-x^*\|_2 \leq  \gamma_2
\left \{2\sigma_k(x)_1+  \left(\|Ax-y\|_1-\varepsilon \right)^+  +
c  (\varepsilon +    \|y-Ax\|_1) \right\},
$$ where $ c $ is the constant given in (\ref{cc1c2}), and  $\gamma _2= \sigma_{\infty, 2} (M^*, M^{**}) $  is the
 Robinson's constant determined by  $ (M^*, M^{**})  $ in $(\ref{M*M**}).$  In particular, for
any $x$ with $\|Ax-y\|_1\leq \varepsilon, $
     there is  an optimal solution $x^*$ of (\ref{l1-l1})
     such that  \begin{equation} \label{DELTA-a} \|x-x^*\|_2 \leq
     2\gamma_2 \{\sigma_k(x)_1 +  c  \varepsilon\}. \end{equation}  }

\emph{Proof.}  Let $x$ be any vector in $\mathbb{R}^n,$ and  let  $(t, r, v_1, \dots , v_5)$ satisfy the following properties:
$t=|x|$, $r=|Ax-y|$, and $(v_1, \dots , v_5)$ is feasible to  (\ref{DLP-A}), i.e.,
$$ A^T (v_3-v_4) + v_1-v_2 =0,  ~ v_1+v_2\leq e, ~  v_3+v_4 \leq v_5 e,  ~ (v_1,\dots, v_5)\geq 0.  $$
For such a vector $(x, t, r, v_1, \dots, v_5)$,   applying Lemma 2.4
with $(M', M'') = (M^*, M^{**}) $ where $M^* $ and $M^{**}$ are the matrices in (\ref{M*M**}),  there exists a point $(x^*,t^*, r^*, v^*_1, \dots ,
  v^*_5)$  in $D^{(1)}
$ defined by (\ref{QQQ}) such that
\begin{equation} \label {hoff-2} \left\|
   \left[
      \begin{array}{c}
        x \\
        t \\
        r \\
        v_1 \\
        \vdots \\
        v_5 \\
      \end{array}\right] - \left[
       \begin{array}{c}
         x^* \\
       t^* \\
         r^* \\
         v^*_1 \\
          \vdots \\
         v^*_5 \\
       \end{array}
     \right]
 \right\|_2 \leq  \gamma_2   \left\| \left[\begin{array}{c}
   (x-t)^+\\
   (-x- t)^+  \\
   (Ax-y-r)^+ \\
 (Ax-y+r)^-\\
 (e^T r-\varepsilon)^+\\
    A^T(v_3-v_4)+v_1-v_2  \\
    ( v_1+v_2-e)^+ \\
    (v_3+v_4-v_5 e)^+\\
      e^T t + y^T (v_3-v_4) +v_5 \varepsilon \\
      ( \vartheta^*) ^-\\
            \end{array}
             \right]\right\|_1,  \end{equation}
             where $(\vartheta^* )^-$ is the short for the vector $((t)^-,
      (r)^-,  (v_1)^-,  \dots, (v_5)^-), $ and
$\gamma_2=\sigma_{\infty, 2}  (M^*, M^{**}) $ is the Robinson's
constant determined  by the matrices $ (M^*, M^{**}) $ in
(\ref{M*M**}).  By the nonnegativity of $ (t,r,v_1, ..., v_5), $  we see that $  (\vartheta^* )^- =0. $ Since $(v_1,
\dots , v_5)$ is feasible to   (\ref{DLP-A}), we also have
$$ (x-t)^+=(-x- t)^+ =0, ~ A^T(v_3-v_4)
+v_1-v_2=0, ~ (v_1+v_2-e)^+=0, $$  $$(v_3+v_4-v_5 e)^+ =0,
  ~ (Ax-y-r)^+=(Ax-y+r)^-=0.
$$
 Thus
     the inequality (\ref{hoff-2}) is reduced to
\begin{eqnarray}
   \|(x, t,  r, v_1,\dots, v_5) -
(x^*,t^*, r^*, v^*_1, \dots ,
  v^*_5) \|_2    \leq    \gamma_2 \left\| \left[
  \begin{array}{c}
     (e^T r-\varepsilon)^+ \\
       e^T t + y^T (v_3-v_4) +v_5 \varepsilon \\
\end{array}
\right] \right\|_1.    \label{S-Hoff}
     \end{eqnarray}
Furthermore, letting $\phi=y -Ax,$ we see that
\begin{eqnarray} \label{eee-2}
 e^T t + y^T (v_3-v_4) +v_5 \varepsilon
 & = & e^T|x|+  (Ax+\phi)^T(v_3-v_4)+   v_5\varepsilon   \nonumber \\
& = & e^T|x|+  x^T A^T  (v_3-v_4) + \phi^T (v_3-v_4)+  v_5\varepsilon.
\end{eqnarray}
Merging (\ref{S-Hoff}) and (\ref{eee-2}) leads to
\begin{eqnarray} \label{INQ-A}
  & &  \|(x, t,  r, v_1,\dots, v_5) -
(x^*,t^*, r^*, v^*_1,\dots,
  v^*_5)\|_2
\nonumber \\
   &    &  \leq     \gamma_2  \left\{(e^T r-\varepsilon)^+  + \left| e^T|x|+  x^T A^T  (v_3-v_4)
   + \phi^T (v_3-v_4)+  v_5\varepsilon  \right| \right\} .
\end{eqnarray}
We now construct a  specific vector
$(\widetilde{v}_1, \dots, \widetilde{v}_5)$  which is feasible to  problem
(\ref{DLP-A}).   To this goal,
 we still let $S$ be  the support set of the $k$-largest components of $|x|,$ and we still decompose $S  $
 as  $S=S_+ \cup S_-, $ where $S_+ =\{i\in S:  x_i> 0\} $
 and $ S_- = \{i\in S: x_i <  0\}.$  Let $\overline{S} =\{1, \dots, n\}\setminus S .$
 Since $A^T$ satisfies the weak RSP of order $k$, there exists a
 vector $ \eta=   A^Tg $ for some
 $g  \in \mathbb{R}^m $  satisfying that
 $ \eta_i= 1\textrm{ for }i\in S_+,  ~ \eta_i= -1 \textrm{ for  }i\in
 S_{-}, $ and $
 |\eta_i| \leq 1\textrm{ for } i \in \overline{S}  .$
Define the vectors $ \widetilde{v}_1 $ and $ \widetilde{v}_2 $ as follows:   $ (\widetilde{v}_1)_i= 1\textrm{ and }
(\widetilde{v}_2)_i=0\textrm{ for }i\in S_+; ~
(\widetilde{v}_1)_i=0\textrm{ and  }(\widetilde{v}_2)_i= 1\textrm{
for }i\in S_-; $ $ (\widetilde{v}_1)_i =(|\eta_i| + \eta_i)/2$ and $
  (\widetilde{v}_2)_i =(|\eta_i| - \eta_i)/2 \textrm{ for all }i\in
\overline{S}. $  This construction ensures that $ (\widetilde{v}_1,
\widetilde{v}_2)\geq 0,  ~ \widetilde{v}_1 + \widetilde{v}_2 \leq e,
$ and $ \widetilde{v}_1-\widetilde{v}_2 =\eta. $
  Moreover, by setting $$\widetilde{v}_3 =|(g)^{-}|=-(g)^{-}, ~ \widetilde{v}_4 =(g)^{+},  ~
     ~ \widetilde{v}_5=\|g\|_\infty,$$ we
  see that $ \widetilde{v}_3 \geq 0,  \widetilde{v}_4\geq 0, $  and
$$   \widetilde{v}_3+\widetilde{v}_4 = -(g)^- + (g)^+  = |g| \leq  \|g\|_\infty e = v_5 e ,   ~ \widetilde{v}_3
-\widetilde{v}_4= -(g)^{-}- (g)^{+} = - g . $$ Note that $A^T
(\widetilde{v}_3 - \widetilde{v}_4) =- A^T g =-\eta
=\widetilde{v}_2-\widetilde{v}_1.$ Therefore, the  vector $
(\widetilde{v}_1,  \dots , \widetilde{v}_5)$ constructed as above is
feasible to the problem (\ref{DLP-A}). We also note that $\eta_S=
\textrm{sign}(x_S)$ and $ \|\eta_{\overline{S}}\|_\infty \leq 1.$
Then it follows from (\ref{INQ-A}) that for the vector $(x, t=|x|,
r=|Ax-y|, \widetilde{v}_1, \dots , \widetilde{v}_5),$ there exists a
point in ${D}^{(1)} ,$ denoted still by $(x^*,t^*, r^*, v^*_1,
\dots,
  v^*_5)$, such that
 \begin{eqnarray} \label{INQ-A3}
  & &  \|(x, t,  r, \widetilde{v}_1, \dots ,\widetilde{ v}_5) -
(x^*,t^*, r^*, v^*_1, \dots,
  v^*_5)\|_2
\nonumber \\
 &  & \leq        \gamma_2  \left[(e^T r-\varepsilon)^+  + \left| e^T|x|+  x^T A^T  (\widetilde{v}_3-\widetilde{v}_4)
   + \phi^T (\widetilde{v}_3-\widetilde{v}_4)+  \widetilde{v}_5\varepsilon  \right|
   \right].  \nonumber \\
 & &  =     \gamma_2  \left[(e^T r-\varepsilon)^+  + \left|\|x\|_1 + (x_S)^T [A^T (\widetilde{v}_3-\widetilde{v}_4)]_S
 +(x_{\overline{S}})^T[A^T  (\widetilde{v}_3-\widetilde{v}_4) ]_{\overline{S}} -
 \phi^Tg+
\widetilde{v}_5 \varepsilon   \right|\right]  \nonumber  \\
 &  &  =     \gamma_2  \left[(e^T r-\varepsilon)^+   + \left|\|x\|_1- \|x_S\|_1   - (x_{\overline{S}})^T\eta_{\overline{S}} -
  \phi^T g +\widetilde{v}_5 \varepsilon
  \right|  \right]  \nonumber  \\
&  & \leq     \gamma_2  \left[(e^T r-\varepsilon)^+ +  \sigma_k(x)_1
+ \|x_{\overline{S}}\|_1 \|\eta_{\overline{S}} \|_\infty + |\phi^Tg|+
\widetilde{v}_5 \varepsilon \right]
   \nonumber  \\
&  & \leq     \gamma_2  \left[(e^T r-\varepsilon)^+ +  2\sigma_k(x)_1
+ \|\phi\|_1 \|g\|_\infty  + \|g\|_\infty \varepsilon \right].
\end{eqnarray}
As $ \| \eta \|_\infty =1 $  and $g=(AA^T)^{-1}A \eta ,$ we have
$\|g\|_\infty \leq \|(AA^T)^{-1}A\|_{\infty \to \infty} =c . $ We
also note that  $r=|Ax-y| =|\phi|, $ which indicates that $e^Tr =
\|Ax-y\|_1 =\|\phi\|_1.$ Thus it follows from (\ref{INQ-A3}) that
 \begin{eqnarray*}   \|x- x^* \|_2
 &    \leq &    \|(x, t, r, \widetilde{v}_1,...,\widetilde{v}_5) -
(x^*,t^*, r^*, v^*_1, ...,
  v^*_5)\|_2
\\
  & \leq &  \gamma_2 \left[ (\|Ax-y\|_1-\varepsilon)^+ +  2\sigma_k(x)_1 +  c  (\|y-Ax\|_1
  +
  \varepsilon) \right].
\end{eqnarray*}
    In particular, when $x$ satisfies the constraint of
(\ref{l1-l1}), i.e., $\|y-Ax\|_1 \leq \varepsilon,$ the above
estimate is reduced to (\ref{DELTA-a}).
The proof is complete. ~  $\Box $

\vskip 0.07in

Similar to Corollary 3.5, we immediately have the following result.

\vskip 0.07in

 \textbf{Corollary 4.5.}   \emph{Let the problem data $(A, y, \varepsilon) $ be given, where $\varepsilon >0, $ $y\in \mathbb{R}^m$ and
 $A\in \mathbb{R}^{m\times n}~  (m<n) $ with $\textrm{rank} (A) =m. $ Let $ c $ and $
    c_1
     $ be the constants given in (\ref{cc1c2}), and let $\gamma_1 $ and $ \gamma_2$ be the
     Robinson's constants given  in Theorems 4.2 and 4.4, respectively.   Suppose that the solutions to (\ref{l1-infty}) and (\ref{l1-l1}) are unique.  If   $A$ satisfies one
of the conditions (p1)--(p6) in Corollary 3.5, then the following
statements hold:}

(i) \emph{For any $x$ satisfying  $\|Ax-y\|_\infty \leq \varepsilon, $
the solution $x^*$ of (\ref{l1-infty}) approximates $x$ with error
 $$ \|x- x^* \|_2 \leq    2 \gamma _1 \{ \sigma_k(x)_1 +  c_1\varepsilon \}.
$$}

 (ii) \emph{For any $x$ satisfying $\|Ax-y\|_1 \leq \varepsilon $, the solution $x^\# $ of (\ref{l1-l1}) approximates $x$ with error
$$ \|x-  x^\# \|_2
   \leq     2 \gamma _2 \{ \sigma_k(x)_1 +  c \varepsilon \}.
$$
}

A difference between Corollary 4.5  and existing
results is in that the constants $\gamma_1$ and
$\gamma_2$  in Corollary 4.5 are Robinson's constants instead of RIP or NSP constants. Each of the matrix
properties  (p1)--(p6) in Corollary 3.5 implies an identical   error
bound.

\section{Robust weak stability of quadratically constrained models}
We now consider the robust weak stability of the
quadratically constrained $\ell_1$-minimization
\begin{equation}\label{l1-l2} \gamma^* : = \min_x \{ \|x\|_1:  ~ \|Ax-y\|_2 \leq \varepsilon
\},
\end{equation}
where $\varepsilon >0$, and $\gamma^*$ denotes the optimal value  of the problem. Let $S^*$   denote the set of optimal
solutions of (\ref{l1-l2}),
which can be represented as
$$ S^* =\{ x\in \mathbb{R}^n : ~ \|x\|_1\leq \gamma^*,  ~ \|Ax-y\|_2\leq
\varepsilon\}. $$   Let  $B =\{z\in \mathbb{R}^m: \|z\|_2\leq 1\}$
be the unit $\ell_2$-ball. Then  problem (\ref{l1-l2}) can be written as
\begin{equation} \label{l1bb} \gamma^*= \min_x \{ \|x\|_1: ~ u=
(Ax-y)/\varepsilon,   ~ u\in B \}.
 \end{equation}
Since the constraint of  (\ref{l1-l2})  is  nonlinear, Lemma 2.4
does not apply to this situation directly. We  need to establish
several auxiliary results in order  to show the robust weak stability of
(\ref{l1-l2}). The main idea is to approximate  $B$ with a polytope.
We recall that $B$ is the intersection of half spaces $a^T z\leq 1$
tangent to its surface, i.e., \begin{equation} \label{Ball} B  = \bigcap_{\|a\|_2=1} \left\{z \in
\mathbb{R}^m: ~ a^T z \leq 1\right\}.
\end{equation}  We also recall  the Hausdorff metric of two sets $S_1, S_2
\subseteq \mathbb{R}^m : $
\begin{equation} \label{HM} \delta^{\cal H} (S_1, S_2) =
\max \left\{\sup_{z'\in S_1} \inf_{z\in S_2} \|z'-z\|_2, ~
\sup_{z\in S_2} \inf_{z'\in S_1} \|z'-z \|_2\right \}.
\end{equation} By taking a finite number of  half-spaces in (\ref{Ball}) to approximate $B, $ Dudley \cite{D74}
 established the following result. (A more discussion
on the polytope approximation of $B$ can be found, for instance, in
\cite{CB00}.)

\vskip 0.07in

 \textbf{Lemma 5.1.}  (Dudley \cite{D74})  \emph{There
exists a constant $ \tau $ such that for every integer number $K >m
$ there is a polytope
\begin{equation} \label{PK} {\cal P}_K= \bigcap_{\|a^i\|_2=1,
 1\leq i\leq K}  \left\{z\in \mathbb{R}^m  : (a^i)^T z\leq 1\right\},\end{equation}   achieving
\begin{equation}  \label{Dudley}  \delta^{\cal H} (B , {\cal P}_K) \leq \frac{\tau}{K^{2/(m-1)}}, \end{equation}  where
$ \delta^{\cal H} (\cdot, \cdot)$ is the Hausdorff metric defined by
(\ref{HM}).}

\vskip 0.07in

From the above lemma, we see that ${\cal P}_K $ can approximate $B$
to any level of accuracy provided that $K$ is sufficiently large.
For ${\cal P}_K$   given by (\ref{PK}), we use   $ M_{{\cal P}_K}:
=[a^1, \dots , a^K]$ to denote the matrix with $a^i\in \mathbb{R}^m, ~i=1, \dots, K$ as its
columns. We also use the symbol $\textrm{Col} (M_{{\cal P}_K})
=\{a^1, a^2, \dots, a^K\} $ to denote the set of columns of
$M_{{\cal P}_K}.$ Thus ${\cal P}_K$ can be written as $$ {\cal P}_K=
\{z\in \mathbb{R}^m: (M_{{\cal P}_K})^T z\leq e\}, $$ where $e$ is
the vector of ones in $\mathbb{R}^K.$  Let $
\{{\cal P}_K \}_{K>m} $ be any sequence of the polytopes given as
(\ref{PK}) and satisfying  (\ref{Dudley}). Consider the sequence of
polytopes $ \{\widetilde{{\cal P}}_J\}_{J>m } ,  $ where
 \begin{equation} \label {PJ-01} \widetilde{{\cal P}}_J = \bigcap_{m< K\leq J} {\cal P}_K  .
\end{equation}   Thus $\widetilde{{\cal P} }_J  $ is still a polytope formed by a
finite number of half space $(a^i) ^T z \leq 1$ where $\|a^i \|_2 =1.$
 We still use $M_{
\widetilde{{\cal P} }_J } $ to denote the matrix with these vectors $a^i$'s as columns,   so $$\widetilde{{\cal P} }_J = \{ z \in \mathbb{R}^m :  (M_{
\widetilde{{\cal P} } _J} )^T  z  \leq e\}.$$
 We still use $ \textrm{Col} (M_{ \widetilde{{\cal P} } _J} ) $  to denote the collection of column vectors of $  M_{
\widetilde{{\cal P} } _J}.$

In what follows, for a given compact convex set $T\subseteq
\mathbb{R}^n$, we denote the projection of $x$ into $T $ by $
\pi_{T} (x) := \textrm{argmin} \{\| x-w\|_2:  ~ w\in T\}. $    We
first prove the following lemma.

\vskip 0.07in

\textbf{Lemma 5.2.}  \emph{Let $\{{\cal P}_K\}_{K>m} $ be any sequence of the polytopes defined by (\ref{PK}) and satisfying  (\ref{Dudley}).  For any $J> m,$  let $\widetilde{{\cal P}
}_J $ be given as (\ref{PJ-01}).  Then for any point $z\in \mathbb{R}^m$ with $\|z\|_2=1,$   there exists a column  vector  $a^i $
of $M_{\widetilde{{\cal P}}_J}, $  i.e., $a^i \in
Col(M_{\widetilde{{\cal P}}_J}),$  such that
 $$ \|z- a^i  \|_2 \leq \sqrt{\frac{2\tau }{J^{2/(m-1)} +\tau } }  .$$}
\emph{Proof.}   Let $z$ be any given point on the unit sphere, i.e.,
$\|z\|_2=1.$
 Since $B \subseteq \widetilde{{\cal P}}_J,  $ where $J >m, $ the straight line passing through  $z$  and the center of $B$ crosses
 a point, denoted by $z',$  on the surface
 of polytope $\widetilde{{\cal P}}_J. $  Clearly,  $z =z'/\|z'\|_2 , $   i.e.,
  $z$ is the projection of $z'$ onto $B. $   Note that $ B \subseteq  \widetilde{{\cal P}}_J \subseteq  {\cal P}_J  $  for any $J >m.$  By the definition of Hausdorff metric and Lemma 5.1, we obtain
 \begin{equation} \label{HD} \|z-z'\|_2 \leq \delta^{\cal H} (B, \widetilde{{\cal P}}_J) \leq \delta^{\cal H} (B, {\cal P}_J) \leq \frac{\tau}{J^{2/(m-1)}}.
 \end{equation}
 Since $z'$ is on the surface of $ \widetilde{{\cal P}}_J,$ there is a vector
 $a^{i_0} \in  \textrm{Col} (M_{ \widetilde{{\cal P}}_J})  $ such that $ (a^{i_0})^T z'=1.$  Note that
 $    \|z'-z\|_2  = \|z'-\frac{z'}{\|z'\|_2}\|_2    =  \|z'\|_2
 -1,
 $
  $\|a^{i_0}\|_2= \|z\|_2 =1$ and $(a^{i_0})^T z'=1 . $ We immediately have
 \begin{eqnarray*}  \|z-a^{i_0}\|_2^2  & =  &  2
 (1-(a^{i_0})^T z) = 2(1-\frac{(a^{i_0})^T z'}{\|z'\|_2})=2(1-\frac{1}{\|z'\|_2}) \\
 & =   &  \frac{2\|z'-z\|_2 } { \|z'-z\|_2 +1}   \leq \frac{2 (\tau /J^{2/(m-1)}) }
  {(\tau /J^{2/(m-1)})    +1} = \frac{2 \tau }  {\tau +J^{2/(m-1)}},  \end{eqnarray*}
 where the inequality follows from (\ref{HD}).   ~ $\Box $

\vskip 0.07in

Recall that  $S^*$ is the set of optimal solutions of (\ref{l1-l2}).
We now prove the next lemma.

\vskip 0.07in

\textbf{Lemma 5.3.} \emph{Let $\{{\cal P}_K\}_{K>m} $ and  $ \widetilde{{\cal P} }_J $ be given
as Lemma 5.2.   Let $ S_{\widetilde{{\cal P} }_J} $ be the set
\begin{equation} \label{S*} S_{{\cal \widetilde{{\cal P} }}_J}    =
\{x \in \mathbb{R}^n:  ~ \|x \|_1\leq \gamma^*,  ~ u=(Ax
-y)/\varepsilon, ~ u\in \widetilde{{\cal P} }_J \}, \end{equation}
where   $\gamma^*$ is the optimal value of (\ref{l1-l2}). Then
$\delta^{\cal H} (S^*, S_{\widetilde{{\cal P} }_J}) \to 0 \textrm{
as } J\to \infty. $}

\vskip 0.07in

\emph{Proof.}   Note that  $
B\subseteq \widetilde{ {\cal P} }_{J} \subseteq {\cal P}_J $ for
every $J > m. $ By the definition of Hausdorff metric and Lemma 5.1, we see that
\begin{equation} \label{BBPP}  \delta^{\cal H} (B, \widetilde{{\cal P} }_{J}) \leq  \delta^{\cal H} (B , {\cal P}_J) \leq
\frac{\tau }{J^{2/(m -1)}}, ~ J >m. \end{equation}    Note
that  $S_{{\cal \widetilde{{\cal P} }}_J},$ given by  (\ref{S*}),  can
be rewritten as  $$ S_{{\cal \widetilde{{\cal P} }}_J}    =   \{x
\in \mathbb{R}^n:  ~ \|x \|_1\leq \gamma^*,  ~ (M_{\widetilde{{\cal
P} }_J})^T (Ax -y) \leq \varepsilon e\},
$$
where $\gamma^*$ is the optimal value of (\ref{l1-l2}). Clearly,
$S^* \subseteq S_ {\widetilde{{\cal P}}_J}$ due to the fact
$B\subseteq  \widetilde{{\cal P}}_J.$ We now prove that $
\delta^{\cal H} (S^*, S_{\widetilde{{\cal P}}_J}) \to 0 $ as $ J \to
\infty. $ Since $S^*$ is a subset of $S_{{\cal \widetilde{P}}_J}$,
by the definition of Hausdorff metric, we see that \begin{equation}
\label{pipi}  \delta^{\cal H} (S^*, S_{{\cal \widetilde{P}}_J}) =
\sup_{w\in S_{{\cal \widetilde{P}}_J}}
 \inf_{z\in S^*}\|w-z\|_2 = \sup_{w\in S_{\widetilde{\cal P}_J}} \| w-\pi_{S^*} (w)\|_2, \end{equation}  where
$\pi_{S^*} (w) \in S^* $ is the projection of
$w$ into $S^*.$  The projection operator $\pi_{S^*} (w) $ is
continuous in $w $ and $S_{\widetilde{\cal P}_J}$ is compact convex set for any $  \widetilde{\cal P}_J .$   Thus for every polytope $\widetilde{{\cal P}}_J,$ the
 superimum in (\ref{pipi})  can be  attained, i.e., there
exists a point, denoted by $w^*_{{\cal \widetilde{P}}_J} \in S_{{\cal
\widetilde{P}}_J},$ such that
 \begin{equation} \label{DDDD}  \delta^{\cal H} (S^*, S_{{\cal \widetilde{P}}_J}) = \left\|w^*_{{\cal \widetilde{P}}_J}  - \pi_{S^*}
 ( w^*_{{\cal \widetilde{P}}_J}) \right\|_2.
\end{equation}
We also note that  $ S^* \subseteq S_{\widetilde{\cal P}_{J+1}}
\subseteq S_{\widetilde{{\cal P}}_J } $ for any $J>m,$ which
implies that $ \delta^{\cal H} (S^*, S_{{\cal \widetilde{P}}_{J+1}})
\leq \delta^{\cal H} (S^*, S_{{\cal \widetilde{P}}_J}). $ Thus
 $ \{\delta^{\cal H} (S^*, S_{\widetilde{{\cal P}}_J})
\}_{J>m} $  is a non-increasing nonnegative sequence. There
must  exist  a number $\delta \geq 0 $ such that
$$ \lim_{J\to \infty} \delta^{\cal H} (S^*, S_{\widetilde{{\cal P}}_J})
=\delta \geq 0. $$ We now further prove that $\delta =0. $  Note
that $ w^*_{{\cal \widetilde{P}}_J} \in S_{\widetilde{{\cal P}}_J}$
for any $J> m.$ Thus
\begin{equation} \label{ggaa} \left\|w^*_{{\cal \widetilde{P}}_J} \right\|_1\leq \gamma^*,  ~ (M_ {{\cal \widetilde{P}}_J})^T (A w^*_{{\cal
\widetilde{P}}_J}- y ) \leq \varepsilon e\textrm{  for any }J >
m. \end{equation}
  The inequality (\ref{ggaa})  implies that the sequence $ \{w^*_{{\cal \widetilde{P}}_J} \} _{J>m} $ is bounded and satisfies that
$$ \left[\sup_{a^i \in \textrm{Col}(M_ {{\cal \widetilde{P}}_J}  ) }   (a^i)^T (Aw^*_{{\cal
\widetilde{P}}_J}-y) \right] \leq \varepsilon  ~ \textrm{  for any  } J
> m .
$$   Note that for any $ m< J' \leq J, $ we have $ \textrm{Col} (M_ {{\cal
\widetilde{P}}_{J'}}   ) \subseteq \textrm{Col} (M_ {{\cal
\widetilde{P}}_J}   ). $ Thus the  inequality above implies that for
any fixed integer number $J' > m,$
  $$ \left[\sup_{a^i \in \textrm{Col}(M_ {{\cal \widetilde{P}}_{J'}})  }  (a^i)^T (Aw^*_{{\cal
\widetilde{P}}_{J}}-y)\right]  \leq \varepsilon ~ \textrm{ for any }J
\geq J'.
$$   Note that the sequence $\{w^*_{{  \widetilde{{\cal P}}_J} } \} _{J\geq J'} $ is bounded.
 Pasting through to a subsequence if necessary, we may assume that $
w^*_{{ \widetilde{{\cal P}}_J }} \to  w^*   $ with $ \|w^*\|_1 \leq
\gamma^*. $  Thus it follows from the  above inequality  that
\begin{equation}\label{JJ''}  \sup_{a^i \in  \textrm{Col}(M_ {{\cal \widetilde{P}}_{J'}}) } (a^i)^T (Aw^*
-y) \leq \varepsilon,  \end{equation}  which holds for any given  $J'
> m.$
 We now prove that (\ref{JJ''}) implies that
$ \|Aw^*  -y\|_2 \leq \varepsilon. $
We show this by contradiction. Assume that  $\|Aw^*  -y\|_2
>\varepsilon, $ which by the definition of the $\ell_2$-norm implies
that
$$ \max _{\|a\|_2=1} a^T(A w^*  -y) = \|A w^*  -y\|_2  >  \varepsilon.$$
 The maximum above attains
at $a^*= (Aw^*  -y)/\|A w^*  -y\|_2. $  By continuity, there exists
a    neighborhood of $a^*, $  namely, $U = a^* + \delta^* B , $
where $\delta^*>0 $ is a small number, such that any point $w\in U
\cap \{z \in \mathbb{R}^m: \|z\|_2=1\} $  satisfies that
\begin{equation} \label{5858}  w^T (Aw^* -y) \geq \frac{1}{2}
\left(\|Aw^* -y \|_2+\varepsilon \right). \end{equation}  Note that
$ \widetilde{{\cal P}}_{J} $ achieves (\ref{BBPP}). Let $J'$  be an
integer number such that $ \sqrt{\frac{2\tau}{(J')^{2/(m-1)}+\tau}}
\leq  \delta^*.$ Applying Lemma 5.2 to $\widetilde{{\cal P}}_{J'},$
we conclude that for the vector $a^*,$  there is a vector $a^i \in
\textrm{Col}(M_ {{\cal \widetilde{P}}_{J'}})  $ such that
$$ \|a^i - a^*\|_2
\leq \sqrt{\frac{2\tau}{(J')^{2/(m-1)}+\tau}} \leq  \delta^*, $$ which, together with the fact $\|a^i\|_2=1, $ implies that
 $a^i \in U \cap \{z \in \mathbb{R}^m: \|z\|_2=1\} . $  Thus it follows from (\ref{5858})  that
$$ (a^i)^T (Aw^* -y) \geq \frac{1}{2} (\|Aw^*  -y \|_2+\varepsilon) >
\varepsilon .
$$ This contradicts  (\ref{JJ''}).
  Thus $w^*$ must satisfy that  $ \|A w^*-y\|_2\leq \varepsilon .$ This together with the fact  $\|w^*\|_1\leq \gamma^*$
  implies that  $w^* \in S^*.$ As a result, $ \pi_{S^*} (w^*)=w^*.$  It follows from (\ref{DDDD}) and
  the continuity of $\pi_{S^*} (\cdot)$
   that
  $$\delta =   \lim_{J\to \infty} \delta^{\cal H} (S^*, S_{\widetilde{{\cal P}}_J}) = \lim_{J\to \infty}
 \| w^*_{{\cal
\widetilde{P}}_J}- \pi_{S^*}(
   w^*_{{\cal
\widetilde{P}}_J})\|_2  =\| w^*-\pi_{S^*} ( w^* )\|_2 =0 ,  $$  as desired.  ~~ $
\Box $

\vskip 0.07in
We will also make use of the following property of a projection operator.
\vskip 0.07in

\textbf{Lemma 5.4.}  \emph{Let $S'$ and $ S''$ be compact convex
sets in $\mathbb{R}^n. $   Then for any $x \in \mathbb{R}^n $, $$
\|\pi_{S'} (x) - \pi_{S''} (x) \|_2^2 \leq \delta^{\cal H} (S',
S'') ( \|x- \pi_{S'} (x) \|_2 + \|x- \pi_{S''} (x) \|_2).
$$}
\emph{Proof.} By the property of projection operators, we have
\begin{equation} \label{Proj-aa} (x- \pi_{S'}(x))^T (v-\pi_{S'}(x)) \leq 0 \textrm{
for all }v \in S',
\end{equation}
\begin{equation} \label{Proj-bb}
 (x- \pi_{S''}(x))^T (u-\pi_{S''}(x)) \leq 0 \textrm{ for all } u \in
S''. \end{equation}    We project  $\pi_{S''}(x) \in S''  $  into
$S'$ to get the point
  $\widehat{v} = \pi_{S'}\left( \pi_{S''}(x)\right)\in S'  $ and we project $\pi_{S'}(x) \in S'  $  into
$S''$ to get the point
  $\widehat{u} = \pi_{S''}\left( \pi_{S'}(x)\right)\in S''.  $  By the definition
  of Hausdorff metric, we have
 \begin{equation} \label{III}  \|\widehat{v} - \pi_{S''}(x)\|_2 \leq \delta^{\cal H}
(S', S''),  ~  \|\widehat{u} - \pi_{S'}(x)\|_2 \leq \delta^{\cal H} (S',
S''). \end{equation}
Substituting $\widehat{v}$ into (\ref{Proj-aa}) and $\widehat{u}$  into
(\ref{Proj-bb}) yields
 $$ (x- \pi_{S'}(x))^T ( \widehat{v}  -\pi_{S'}(x)) \leq 0, ~ (x- \pi_{S''}(x))^T (\widehat{ u} -\pi_{S''}(x)) \leq 0, $$ which
 implies the first inequality below \begin{eqnarray*}
  \|\pi_{S'}(x) -\pi_{S''}(x)\|_2^2  & = & (\pi_{S'}(x) -x+x-\pi_{S''}(x))^T (  \pi_{S'}(x) -\pi_{S''}(x) )\\
   & = &
 -(x- \pi_{S'}(x))^T ( \pi_{S'}(x)
 -\pi_{S''}(x)) +  (x- \pi_{S''}(x))^T ( \pi_{S'}(x)
 -\pi_{S''}(x)) \\
 &\leq  &  - (x- \pi_{S'}(x))^T ( \widehat{v}- \pi_{S''}(x)
 ) +  (x- \pi_{S''}(x))^T ( \pi_{S'}(x)- \widehat{u}) \\
 & \leq & \| x- \pi_{S'}(x)\|_2 \| \pi_{S''}(x)- \widehat{v}\|_2  + \| x- \pi_{S''}(x)\|_2 \|
  \pi_{S'}(x)- \widehat{u} \|_2  \\
 & \leq & \delta^{\cal H}
(S', S'')  (\| x- \pi_{S'}(x)\|_2 +  \| x- \pi_{S''}(x)\|_2),
 \end{eqnarray*}
where the final  inequality   follows from   (\ref{III}). ~  $\Box$

\vskip 0.07in

For each $K>2m,$ by Lemma 5.1, there is a polytope ${\cal P}_K$ of
the form (\ref{PK})   achieving (\ref{Dudley}), and ${\cal P}_K $ can
be represented as   ${\cal P}_K=  \{ z \in
\mathbb{R}^m: (M_{{\cal P}_K})^T z\leq e\} . $  We now add the following $2m$
half spaces $$(\pm \varrho_i)^T z \leq 1, ~i=1, \dots, m $$ to ${\cal
P}_K $,   where $\varrho_i  ~(i=1, \dots , m) $  denotes the  $i$-th column vector of the $m\times m$ identity
matrix. Let $\widehat{K}$ denote the cardinality of the set
$\textrm{Col} (M_{ {\cal P}_ K } ) \cup \{\pm \varrho_i: i=1, \dots,
m\}. $ This yields the polytope
\begin{equation}  \label {PK-C} {\cal P}_{\widehat{K}} : = {\cal P}_K \cap \{z\in \mathbb{R}^m:   ~ \varrho_i^T z\leq 1,
~ -\varrho_i^Tz \leq  1, ~ i=1, \dots, m\},
 \end{equation} Therefore,
\begin{equation} \label{col-col} \textrm{Col} (M_{ {\cal P}_{\widehat{K}}
}) = \textrm{Col} (M_{ {\cal P}_K } ) \cup \{\pm \varrho_i: i=1,
\dots, m\}
\end{equation} and
  $ \widehat{K} =  |\textrm{Col} (M_{ {\cal P}_{\widehat{K}} })|. $   Clearly, $ K \leq \widehat{K} \leq K+2m$
which together with $K>2m $ implies that $ 1 \leq \widehat{K} /K\leq 2.$
Let $\tau$ be the constant in Lemma 5.1 and let
$\tau' = 4^{1/(m-1)}\tau. $  By the definition of Hausdorff metric
and Lemma 5.1, we see that the polytope ${\cal P}_{\widehat{K}}$
constructed as (\ref{PK-C}) satisfies
\begin{equation} \label{6868}  \delta^{\cal H} (B, {\cal P}_ {\widehat{K}} ) \leq
\delta^{\cal H} (B, {\cal P}_K) \leq \frac{\tau} {K^{2/(m-1)}} =
\frac{\tau} {\widehat{K}^{2/(m-1)}} \left(\frac {\widehat{K}}{K}\right)
^{2/(m-1)} \leq \frac{\tau'} {\widehat{K}^{2/(m-1)}}.
\end{equation}
We use the set ${\cal P}_{\widehat{K}}$ defined as
(\ref{PK-C}), which achieves (\ref{6868}), to construct the sequence of polytopes $\{\widetilde{{\cal P}}_J\}$   as follows:
\begin{equation} \label{PJ} \widetilde{{\cal P}}_J  = \bigcap_{m< \widehat{K}
\leq J } {\cal P}_{\widehat{K}} .
\end{equation}  Let $S_{\widetilde{{\cal P}}_J} $  denote the set
  (\ref{S*}) with $ \widetilde{{\cal P}}_J  $ being given by (\ref{PJ}).  Then  Lemma 5.3 remains valid for the sequence of polytopes given by (\ref{PJ}). So $\delta^{\cal H} (S^*,
S_{\widetilde{{\cal P}}_J}) \to 0$  as $ J\to \infty.$

 Thus in the remainder of the paper, let $\varepsilon' >0$ be
any fixed small number. From the above discussion,  there exists an integer number $J_0
> 2m$ such that
\begin{equation} \label{var-var} \delta^{{\cal H}} (S^*, S_{\widetilde{{\cal
P}}_{J_0}})  \leq \varepsilon'.
\end{equation}
We consider the fixed polytope $\widetilde{{\cal
P}}_{J_0}$ constructed as above. This polytope is an approximation of $B  $ and
achieves (\ref{var-var}).   We use
$\widehat{N}$ to denote the number of columns of
$M_{\widetilde{{\cal P}}_{J_0}}$ and use  $e_{\widehat{N}}$ to
denote the vector of ones in $\mathbb{R}^{\widehat{N}}$ to
distinguish it from $e,$ the vector of ones  in $\mathbb{R}^n. $
Replacing $B $ in  (\ref{l1bb}) by $\widetilde{{\cal P}} _ {J_0}  $
leads to the following approximation of (\ref{l1-l2}):
\begin{equation} \label {PPJJ} \gamma^*_{\widetilde{{\cal P} }_{J_0}}: =\min_x\{\|x\|_1: ~ u=(Ax-y)/\varepsilon, u\in
\widetilde{{\cal P} }_{J_0}\} = \min_x\{\|x\|_1: ~
(M_{\widetilde{{\cal P}} _{J_0}} )^T (Ax-y) \leq \varepsilon
e_{\widehat{N}}\} , \end{equation} where $\gamma^*_{\widetilde{{\cal
P} }_{J_0}} $ is the optimal value of the above problem.  Let $$
S^*_{\widetilde{{\cal P} }_{J_0}}  =  \{ x\in \mathbb{R}^n : ~
\|x\|_1 \leq \gamma^*_{\widetilde{{\cal P} }_{J_0}}, ~
u=(Ax-y)/\varepsilon,  ~ u\in
 \widetilde{{\cal P}}_{J_0} \} $$ be the set of optimal solutions of  (\ref{PPJJ}), and  let $S_ {{\cal
\widetilde{P}}_{J_0} } $ be the set defined by (\ref{S*}) with $\widetilde{{\cal P}}_J$ replaced by $\widetilde{{\cal P}}_{J_0} .$    Clearly,
$ S^* \subseteq   S_{{\cal \widetilde{P}}_{J_0} }. $ Note that $
\gamma^*_{\widetilde{{\cal P} }_{J_0}}  \leq \gamma^*   $ due to the
fact $B\subseteq \widetilde{{\cal P} }_{J_0}. $  We immediately see
that $S^*_{\widetilde{{\cal P}}_{J_0}} \subseteq S_{{\cal
\widetilde{P}}_{J_0} }  $ The  problem  (\ref{PPJJ})  can be written
as
$$  \min_{(x,t)} \{ e^T t: ~ x\leq t, ~ -x\leq t, ~ t\geq 0, ~(M_{\widetilde{{\cal P}}_{J_0}})^T (Ax-y) \leq
\varepsilon e_{\widehat{N}}\}, $$ to which the dual problem is given as
\begin{eqnarray} \label{ddll} &  \max  &  - \left[\varepsilon e_{\widehat{N}} + (M_{\widetilde{{\cal
P}}_{J_0}})^Ty \right]^T v_3   \\
& \textrm{s.t.} &  A^T M_{\widetilde{{\cal P}}_{J_0}} v_3
+v_1-v_2=0, ~ v_1+v_2 \leq e, ~(v_1, v_2, v_3)\geq 0.  \nonumber
\end{eqnarray}

The following lemma immediately follows from  the optimality
condition of the above linear program.

\vskip 0.07in

\textbf{Lemma 5.5.} \emph{$x^* \in \mathbb{R}^n$ is an optimal
solution of (\ref{PPJJ})  if and only if there exist vectors $t^*,
v_1^*, v_2^* \in \mathbb{R}^n_+ $ and $
 v_3^* \in \mathbb{R}^{\widehat{N}} _+  $   such that $ (x^*, t^*,
v_1^*, v_2^*, v_3^*) \in {D}^{(2)} $ where
 \begin{equation} \label{TTT} \begin{array} {ll}  {D}^{(2)} =\{(x,t,  v_1,v_2, v_3):
 & x\leq t, ~ -x\leq t, ~  (M_{\widetilde{{\cal P} }_{J_0}})^T (Ax-y)\leq \varepsilon e_{\widehat{N}},
      \\
 &  A^T  M_{\widetilde{{\cal P} }_{J_0}} v_3 + v_1-v_2 =0,
   ~ v_1+v_2\leq e,    \\
&  e^T t =- \left[\varepsilon e_{\widehat{N}}+ (M_{\widetilde{{\cal P}}_{J_0}})^T y \right] ^T v_3,
~  (t,  v_1, v_2,   v_3) \geq 0\}.
\end{array}  \end{equation}
 Moreover, for any $ (x,t, v_1,v_2,  v_3) \in D^{(2)}, $ it must hold that
 $t=|x|.$ }

\vskip 0.07in
 To apply Lemma 2.4, we write (\ref{TTT}) in the
form
 \begin{equation} \label{TTT2} {D}^{(2)}  = \{z= (x,t, v_1,v_2,  v_3):   ~ M^+ z\leq b^+,  ~ M^{++}  z = b^{++} \},
 \end{equation}  where $b^{++}=0$  and
 \begin{equation} \label {M+b+}  M^+ = \left(
            \begin{array}{ccccc}
              I & -I & 0 & 0 & 0 \\
              -I & -I & 0 & 0 & 0 \\
              (M_{\widetilde{{\cal P}}_{J_0}})^T A & 0 & 0 & 0 & 0 \\
              0 & 0 & I & I & 0 \\
              0 & -I & 0 & 0 & 0 \\
              0 & 0 & -I & 0 & 0 \\
              0 & 0 & 0 & -I & 0 \\
              0 & 0 & 0 & 0 & -I_{\widehat{N}} \\
            \end{array}
          \right)
 , ~~ b^+ =  \left(
            \begin{array}{c}
              0 \\
              0 \\
              (M_{\widetilde{{\cal P}}_{J_0}})^Ty +\varepsilon e_{  \widehat{N}} \\
              e\\
              0 \\
              0 \\
              0 \\
              0 \\
            \end{array}
          \right),
  \end{equation}
\begin{equation} \label{M++}  M^{++} =  \left(
                \begin{array}{ccccc}
                  0 &  0 & I  & -I  & A^T M_{\widetilde{{\cal P}}_{J_0}} \\
                  0 & e^T & 0  & 0  &  \varepsilon e_{\widehat{N}} ^T +   y^T M_{\widetilde{{\cal P}}_{J_0}} \\
                \end{array}
              \right),
\end{equation}
where $I$ and $I_{\widehat{N}}$ are the $n\times n$  and $\widehat{N}\times \widehat{N}$ identity
matrices, respectively.
We now prove the main result in this section.

 \vskip 0.07in

\textbf{Theorem 5.6.}    \emph{Let the problem data $(A, y,
\varepsilon)$  of (\ref{l1-l2}) be given, where $\varepsilon >0,$
$y\in \mathbb{R}^m$  and $ A \in \mathbb{R}^{m\times n} ~(m<n) $ with
$\textrm{rank}(A)=m. $  Let $\varepsilon'$ be any prescribed small number and let the polytope $\widetilde{{\cal
P}}_{J_0}$ be constructed as (\ref{PJ}) and
achieve (\ref{var-var}).   Suppose that $A^T$ satisfies the weak RSP of order
$k. $
   Then for any $x \in \mathbb{R}^n, $
     there is an optimal solution $x^*$ of (\ref{l1-l2})
     such that \begin{equation}\label{l2-error} \|x-x^*\|_2 \leq
     2 \gamma_3
\left\{ \widehat{N}(\|Ax-y\|_2-\varepsilon)^+   +2\sigma_k(x)_1 + c_1
\varepsilon  + c_2 \|Ax-y\|_2 \right\} + 2\varepsilon', \end{equation}
where $ c_1
 $ and $  c_2 $ are constants given in (\ref{cc1c2}),    $\gamma _3 = \sigma_{\infty, 2} (M^+, M^{++}) $ is
 the Robinson's
  constant determined by  $(M^+, M^{++})$ given in (\ref{M+b+}) and  (\ref{M++}).
 Moreover,  for any $x$ with $\|Ax-y\|_2\leq \varepsilon, $
      there is an optimal solution $x^*$ of (\ref{l1-l2})
     such that  $$ \|x-x^*\|_2 \leq
     4\gamma_3 \sigma_k(x)_1 + 2 \gamma_3  (c_1+  c_2)
     \varepsilon + 2 \varepsilon'.
     $$}
\emph{Proof.}  Let $x$ be any vector in $\mathbb{R}^n$ and let
$t=|x|.$ We still denote by $S$ the support set of the $k$-largest
entries of $|x|.$  Let $S_+ =\{ i\in S: x_i> 0\} $
 and $ S_- = \{i\in S: x_i <  0\} .$ Then  $S=S_+ \cup S_-. $
 Since $A^T$ satisfies the weak RSP
 of order $k$, there exists a vector  $ \eta= A^Tg  $ for some
 $g\in \mathbb{R}^m $,  satisfying that
 $ \eta_i= 1 \textrm{ for }i\in S_+,  ~ \eta_i= -1 \textrm{ for  }i\in S_{-}, \textrm{ and }
 |\eta_i| \leq 1\textrm{ for } i \in \overline{S} , $ where $\overline{S} =\{
 1, \dots, n\} \backslash S. $
For the given problem data $(A, y, \varepsilon),$ as shown between (\ref
{PK-C}) and (\ref{var-var}), there exists an integer number $J_0 > 2m $
such that the polytope $\widetilde{\mathcal{P}}_{J_0},$  given as (\ref{PJ}),
can approximate $B$ and
 achieve  the  bound (\ref{var-var}).   We now construct a
feasible solution $(\widetilde{v}_1, \widetilde{v}_2,
\widetilde{v}_3)$ to  problem (\ref{ddll}). Set
$(\widetilde{v}_1)_i= 1\textrm{ and } (\widetilde{v}_2)_i=0\textrm{
for all }i\in S_+, ~ (\widetilde{v}_1)_i=0\textrm{ and
}(\widetilde{v}_2)_i= 1\textrm{ for all }i\in S_- , $ and  $
(\widetilde{v}_1)_i =(|\eta_i|+\eta_i)/2 $ and $  (\widetilde{v}_2)_i
=(|\eta_i|-\eta_i)/2 \textrm{ for all }i\in \overline{S}. $ This
 choice of $\widetilde{v}_1$ and $\widetilde{v}_2$ ensures that $ (\widetilde{v}_1, \widetilde{v}_2)\geq
0,  ~  \widetilde{v}_1 + \widetilde{v}_2 \leq e $ and $
\widetilde{v}_1 -\widetilde{v}_2  =\eta. $ We now construct the
vector $\widetilde{v}_3. $
    By the construction of $\widetilde{{\cal P}}_{J_0}$, we see that
    $$ \{\pm \varrho_i: ~i=1, \dots,  m\} \subseteq \textrm{Col}  (M_{\widetilde{{\cal
    P}}_{J_0}}).
    $$
   It is not difficult to show that
    there exists a
vector $\widetilde{v}_3\in \mathbb{R}^{\widehat{N}}_{+}$ satisfying
$M_{\widetilde{{\cal P}}_{J_0} } \widetilde{v}_3= - g $ and $
\|\widetilde{v}_3\|_1= \|g\|_1.$  In fact, without
loss of generality, we assume that $ \{-\varrho_i: ~i=1, \dots, m\}$
are arranged as the first $m$ columns  and  $ \{\varrho_i: ~i=1,
\dots,  m\}$ are arranged as the second $m$ columns in
$M_{\widetilde{{\cal P}}_{J_0} }. $ For every $i=1, \dots , m,$ if
$g_i\geq 0,$ then we set  $ (\widetilde{v}_3)_i= g_i; $ otherwise,
if $g_i< 0,$ then we set   $ (\widetilde{v}_3)_{m+i}=  - g_i.  $ All
remaining entries of $ \widetilde{v}_3  \in \mathbb{R}^{\widehat{N}}
$ are set to be zero. By this choice of $ \widetilde{v}_3,$ we see
that $\widetilde{v}_3 \geq 0$, $M_{\widetilde{{\cal P}}_{J_0} }
\widetilde{v}_3= - g  $ and
\begin{equation} \label{E1}  \|\widetilde{v}_3\|_1 = \|g\|_1 = \|(AA^T)^{-1}A \eta\|_1\leq
\|(AA^T)^{-1}A \|_{\infty\to 1} \|\eta\|_\infty \leq  c_1,
\end{equation}
 where $  c_1  $ is the constant given in (\ref{cc1c2}).

 Let $ D^{(2)} $ be given as  in Lemma 5.5.  $ D^{(2)} $ can be written as (\ref{TTT2}).
 For the vector $(x,t, \widetilde{v}_1, \widetilde{v}_2, \widetilde{v}_3), $ applying Lemma 2.4 with $(M', M'')= (M^+, M^{++}) $ where $M^+$ and $ M^{++}$ are given as  (\ref{M+b+}) and (\ref{M++}),
  there exists a point in $ D^{(2)} ,$ denoted by
$(\widehat{x}, \widehat{t}, \widehat{v}_1, \widehat{v}_2,
\widehat{v}_3), $ such that
\begin{equation} \label {hoff-4} \left\|
   \left[
      \begin{array}{c}
        x \\
        t \\
        \widetilde{v}_1 \\
        \widetilde{v}_2 \\
        \widetilde{v}_3 \\
      \end{array}\right] - \left[
       \begin{array}{c}
         \widehat{x} \\
        \widehat{t} \\
         \widehat{v}_1 \\
         \widehat{v}_2 \\
         \widehat{v}_3 \\
       \end{array}
     \right]
 \right\|_2 \leq  \gamma_3 \left\| \left[\begin{array}{c}
    \left((M_{\widetilde{{\cal
P}}_{J_0} })^T(Ax-y)-\varepsilon e_{ \widehat{N} } \right)^+ \\
(x-t)^+ \\
   (-x- t)^+ \\
    A^T M_{\widetilde{{\cal
P}}_{J_0} } \widetilde{v}_3+ \widetilde{v}_1-\widetilde{v}_2  \\
    ( \widetilde{v}_1+\widetilde{v}_2-e)^+ \\
      e^T t + \left(\varepsilon e_{ \widehat{N} } +  (M_{\widetilde{{\cal
P}}_{J_0} })^T y \right) ^T\widetilde{v}_3 \\
      t^-\\
        (\widetilde{\vartheta})^-
            \end{array}
             \right] \right\|_1, \end{equation}
    where  $(\widetilde{\vartheta})^-$ denotes the vector $
( (\widetilde{v}_1)^-,
          (\widetilde{v}_2)^-,
          (\widetilde{v}_3)^-),
      $ and  $\gamma_3 =\sigma_{\infty, 2} (M^+, M^{++}) $ is the Robinson's constant
    determined by  $(M^+, M^{++})$ given in (\ref{M+b+}) and (\ref{M++}).
Note that   $t=|x|$ implies that  $(x-t)^+ =
   (-x- t)^+  =t^- =0.  $ Also, since  $ (\widetilde{v}_1,
\widetilde{v}_2, \widetilde{v}_3) $ is feasible to (\ref{ddll}), we
have   $ (\widetilde{\vartheta})^- =0,$  $( \widetilde{v}_1+\widetilde{v}_2-e)^+ =0$ and $A^T M_{\widetilde{{\cal
P}}_{J_0} } \widetilde{v}_3+ \widetilde{v}_1-\widetilde{v}_2 =0.$
Thus   (\ref{hoff-4}) is reduced to
\begin{equation}  \label {error-A} \|x-\widehat{x} \|_2 \leq \gamma_3 \left\{ \left\|
\left[(M_{\widetilde{{\cal P}}_{J_0} })^T(Ax-y)-\varepsilon e_N \right]^+ \right\|_1+
\left|e^T t +  \left[\varepsilon e_{\widehat{N}} + (M_{\widetilde{{\cal P}}_{J_0}
})^T y  \right]^T \widetilde{v}_3\right|\right\}.
\end{equation}
 Note
that for every $a^i \in \textrm{Col}(
M_{\widetilde{{\cal P}}_{J_0} }),$ we have  $\|a^i\|_2 =1 $  and thus $ (a^i)^T (Ax-y)\leq
\|Ax-y\|_2.$ This implies that $ \left[(a^i)^T (Ax-y)- \varepsilon\right]^+ \leq
\left(\|Ax-y\|_2-\varepsilon\right)^+  $ and hence
$$ \left[(
 M_{\widetilde{{\cal P}}_{J_0} })^T  (Ax-y) -\varepsilon e_{\widehat{N}} \right]^+ \leq (\|Ax-y\|_2
-\varepsilon)^+ e_{\widehat{N}}, $$ and hence
\begin{equation} \label{NNNN}  \left\|\left[(M_{\widetilde{{\cal P}}_{J_0} })^T(Ax-y)-\varepsilon e_{\widehat{N}} \right]^+\right\|_1 \leq \widehat{N}
(\|Ax-y\|_2-\varepsilon  )^+. \end{equation}   By the definition of $\eta, $  we
see that  $x^T A^T g= x^T \eta  = \|x_S\|_1+ x_{\overline{S}}^T
\eta_{\overline{S}}  $ and  thus
$$ \left|e^T |x| - x^T A^T g\right|= \left|\|x\|_1-\|x_S\|_1 - x_{\overline{S}}^T
\eta_{\overline{S}} \right|\leq \| x_{\overline{S}}\|_1
+|x_{\overline{S}}^T \eta_{\overline{S}}|\leq 2
\|x_{\overline{S}}\|_1 =2\sigma_k(x)_1.$$ We also note that
\begin{equation} \label{E2}  \|g\|_2  = \|(AA^T)^{-1}A \eta\|_2 \leq
\|(AA^T)^{-1}A \|_{\infty\to 2} \|\eta\|_\infty \leq  c_2,
\end{equation} where $  c_2  $ is the constant given in (\ref{cc1c2}).
 Thus, by letting  $\phi= Ax-y$ and noting that  $M_{\widetilde{{\cal P}}_{J_0} }\widetilde{v}_3=-g,$   we have
 \begin{eqnarray} \label {error-B}  \left|e^T t +\left[\varepsilon e_{\widehat{N}} + (M_{\widetilde{{\cal P}}_{J_0} })^T y
\right]^T \widetilde{v}_3 \right|
& = & \left|e^T |x|   +x^T A^T M_{\widetilde{{\cal P}}_{J_0} }
\widetilde{v}_3- \phi^T  M_{\widetilde{{\cal P}}_{J_0} } \widetilde{v}_3 + \varepsilon e^T_{\widehat{N}} \widetilde{v}_3 \right|  \nonumber \\
& = & |e^T |x| - x^T A^T g + \phi^T g  + \varepsilon e^T_{\widehat{N}} \widetilde{v}_3 | \nonumber \\
& \leq  & 2\sigma_k(x)_1  + |\phi^T
g| + |\varepsilon e^T_{\widehat{N}} \widetilde{v}_3|   \nonumber \\
& \leq & 2\sigma_k(x)_1  +
\|\phi\|_2
\| g \|_2 + \varepsilon \|\widetilde{v}_3\|_1 \nonumber \\
& \leq & 2\sigma_k(x)_1  + c_2 \|Ax-y\|_2+  \varepsilon c_1,
\end{eqnarray}
where the last inequality follows from (\ref{E1}) and (\ref{E2}).
Merging (\ref{error-A}), (\ref{NNNN}) and  (\ref{error-B}) leads to
\begin{equation}\label{error-C} \|x-\widehat{x}\|_2 \leq
     \gamma_3
\left[ |{\widehat{N}}|(\|Ax-y\|_2-\varepsilon)^+   +2\sigma_k(x)_1 + c_1
\varepsilon  + c_2 \|Ax-y\|_2 \right]. \end{equation}
 Note that the set $S_{\widetilde{{\cal P}}_{J_0}}  $ and $S^*$
  are compact convex sets. Let
$x^*$ and $ \overline{x} $ denote the  projection of $x$
onto $S^*$ and $S_{ \widetilde{{\cal P}}_{J_0} }$ respectively,
namely, $ x^* = \pi_{S^*} (x)  \in S^* $ and $ \overline{x}  =
\pi_{S_{\widetilde{{\cal P}}_{J_0}} } (x) \in
S_{\widetilde{{\cal P}}_{J_0}}.  $  Since $S^* \subseteq S_{ \widetilde{{\cal P}}_{J_0} }, $ we have $\|x-\overline{x}\|_2\leq \|x-x^*\|_2. $   By
(\ref{var-var}),   $ \delta^{\cal H} (S^*, S_{\widetilde{{\cal
P}}_{J_0} })  \leq \varepsilon', $ which together with Lemma 5.4
implies that
\begin{equation} \label{IIFF} \|x^*- \overline{x} \|_2 ^2 \leq \delta^{\cal H} (S^*,  S_{\widetilde{{\cal P}}_{J_0} })
  (\|x-x^*\|_2 + \|x-\overline{x} \|_2)  \leq  \varepsilon'  (\|x-x^*\|_2 + \|x-\overline{x} \|_2) . \end{equation}
 Note that $\widehat{x} \in S^*_{\widetilde{{\cal P}}_{J_0} } \subseteq  S_{\widetilde{{\cal P}}_{J_0} } $  and $\overline{x} $ is the projection of $x$ into the convex set
  $S_{\widetilde{{\cal P}}_{J_0} }. $ Thus $ \|x-\overline{x} \|_2 \leq  \|x-\widehat{x} \|_2 .$
By triangle inequality and (\ref{IIFF}), we have \begin{eqnarray} \label{error-D}
\|x-x^*\|_2   & \leq & \|x-\overline{x} \|_2+ \| \overline{x} -x^*\|_2  \nonumber \\
 & \leq &  \|x-\widehat{x}\|_2 +  \|\overline{x} - x^* \|_2  \nonumber \\
 & \leq & \|x-\widehat{x}\|_2 + \sqrt{\varepsilon' (\|x-x^*\|_2 + \|x-\overline{x} \|_2) . }
 \end{eqnarray}
Since $ \|x-\overline{x} \|_2 \leq \|x-x^*\|_2 , $   it
follows from (\ref{error-D}) that
\begin{eqnarray} \label{error-E}
\|x-x^*\|_2   \leq  \|x-\widehat{x}\|_2 + \sqrt{2\varepsilon'
\|x-x^*\|_2  },
 \end{eqnarray}
 which implies
that
 $$ \|x-x^*\|_2  \leq  \left(\frac{\sqrt{2\varepsilon'} +\sqrt{2\varepsilon'+4\|x-\widehat{x}\|_2 }}{2}  \right)^2
 \leq 2\varepsilon' + 2 \|x-\widehat{x}\|_2,  $$ where the last inequality follows from the fact $\left(\frac{a+b}{2}\right)^2 \leq \frac{a^2+b^2}{2}.$
Combination of the inequality above and (\ref{error-C}) immediately yields (\ref{l2-error}), i.e.,
$$ \|x-x^*\|_2 \leq 2\varepsilon' + 2\gamma_3 \left\{\widehat{N} (\|Ax-y\|_2-\varepsilon  )^+ +  2\sigma_k (x)_1 +
c_1 \varepsilon + c_2\|Ax-y\|_2 \right\}. $$   In particularly, when $x $ satisfies
$\|Ax-y\|_2\leq \varepsilon$, the above inequality is reduced to
$$ \|x-x^*\|_2 \leq 2\varepsilon' + 2\gamma_3 \left\{ 2\sigma_k(x)_1 +
(c_1+c_2) \varepsilon\right\} =4\gamma_3  \sigma_k(x)_1 +  2\gamma_3(c_1+c_2)\varepsilon +2\varepsilon', $$  as desired.  ~~ $ \Box $

\vskip 0.07in
We immediately have the following corollary.

\vskip 0.07in

\textbf{ Corollary 5.7.} \emph{Let the problem data $(A, y,
\varepsilon)$  be given, where $\varepsilon >0$, $y\in \mathbb{R}^m$
and $ A \in \mathbb{R}^ {m\times n}  (m<n )$ with $rank (A) =m.$  Let $\varepsilon'$ be any prescribed small number and the polytope $\widetilde{{\cal
P}}_{J_0}$
achieve (\ref{var-var}). Then
under each of the listed conditions in Corollary 3.5, for any $x\in
\mathbb{R}^n$ with $\|Ax-y\|_2 \leq \varepsilon $ there is an
optimal solution $x^*$ of (\ref{l1-l2})
     such that
$$ \|x-x^*\|_2 \leq
 4\gamma_3  \sigma_k(x)_1 +  2(\gamma_3 c_1+ \gamma_3c_2 )  \varepsilon +2 \varepsilon' .$$
where  $ c_1 $ and $ c_2$ are given in (\ref{cc1c2}) and $ \gamma_3$ is the Robinson's constant given  in Theorem 5.6.}

 \vskip 0.07in

 The weak stability is a more general concept than stability. Any traditional sufficient condition for stability of $\ell_1$-minimization problems, by Theorem 2.3,    implies the weak RSP of $A^T.$ From a mathematical point of view, we have completely characterized the weak stability of standard $\ell_1$-minimization  under this assumption (see Corollary 3.3). It is worth emphasizing several important features of the weak RSP of $A^T. $

 (\textbf{i})  Uniform recovery of every $k$-sparse vector is a basic requirement in compressed sensing, and the classic KKT optimality condition is a fundamental tool for understanding the internal mechanism of $\ell_1$-minimization methods.    The weak RSP of $A^T$ is a natural property capturing both the requirement of uniform recovery and the deepest property of any optimal solution to $\ell_1$-minimization.  So our assumption is actually a strengthened KKK optimality conditions by taking into account the requirement of uniform recovery. As a result, no matter what (deterministic or random)  matrix $A$ is used, the weak RSP of $A^T$ is a fundamental property  guaranteeing the success and stableness of $\ell_1$-minimization methods in sparse data recovery. As shown by Corollary 3.3, this property cannot be relaxed without damaging the weak stability of $\ell_1$-minimization, since it is a necessary and sufficient condition for $\ell_1$-minimization   to be weakly stable for any measurement  $y\in \{Ax: \|x\|_0\leq k\}$.

 (\textbf{ii})  Our analysis is different from the existing frameworks. It is based on the Hoffman's error bound  for linear systems
 and the polytope approximation of the unit $\ell_2$-ball. The weak RSP of $A^T$ is a constant-free matrix property. The    coefficients  $C, C_1 $ and $C_2$  in error bounds (\ref{S}) and
(\ref{RS}) are measured by the Robinson's constants, no matter the matrix property is  constant-free  (such as the weak RSP of $A^T$,   RSP of order $k$ of $A^T$, or NSP of order $k$) or is constant-dependent  (such as the RIP, stable or robust stable NSP). Thus our analytic method yields a certain unified weak stability result irrespective of an individual assumption on $A$, provided that the imposed assumption implies the weak RSP of $A^T $ (see Corollaries 3.5, 4.5 and 5.7).

(\textbf{iii})  Practical signals are often structured or with some prior information, and typical design matrices in practice are not   Gaussian or Bernoulli. This makes the standard analysis and  results (based on Gaussian and Bernoulli random matrices) difficult to apply   in these situations. Thus the structured sparse data reconstruction recently becomes one of the active research areas in compressed sensing and applied mathematics.  The weak RSP concept  derived from optimality conditions of convex optimization  can be easily adapted to these situations to interpret the behavior of more complex and  general recovery problems.   For instance, the so-called restricted RSP property of $A^T$  was used to deal with the sign  or support  recovery of signals in 1-bit compressed sensing problems \cite{ZX16}.

It is also worth mentioning that the analytic method  in this
paper is not difficult to be extended to the study of the weak stability of  weighted
$\ell_1$-minimization problems (e.g., \cite{CWB08, ZL12, ZK15}), Dantzig selector \cite{CT07},  and Lasso problems \cite{T96, HTW15}.

\section{Conclusions}

We have shown that the so-called weak range space property of the
transposed design matrix  is a sufficient
 constant-free condition for various $\ell_1$-minimization problems to be (robustly and) weakly stable in sparse data reconstruction.
  For noise-free measurements, this matrix property  turns out to be a necessary condition
for   standard $\ell_1$-minimization to be weakly stable. All existing
stability conditions (such as mutual coherence, RIP, NSP, or their
variants) imply  our assumption. As a result,  certain  unified weak stability results have been developed for $\ell_1$-minimization   under existing matrix properties. In particular, the weak stability under the constant-free null space property of order $k$  and range space property of order $k$ have been established in this paper.   Our  stability coefficients
  are measured by the  Robinson's
constants  determined by the  problem data.
 Our study indicates that the reconstruction error bounds via $\ell_1$-minimization
   can be understood from   Hoffman's
error bounds for  linear systems with
 compressed sensing matrices.

\end{document}